\documentclass[11pt,a4paper]{article}

\pdfoutput=1
\usepackage{amsmath,amsfonts,amsbsy,amssymb,enumerate,array}
\usepackage{graphicx}
\usepackage[usenames]{color}
\usepackage[english]{babel}
\usepackage{fancybox}
\usepackage{slashed}
\usepackage{chngpage} 

\definecolor{AV}{rgb}{0.65,0.0,0}
\definecolor{DT}{rgb}{0,0,0.65}

\newcommand{\captionfonts}{\small}

\makeatletter  
\long\def\@makecaption#1#2{%
  \vskip\abovecaptionskip
  \sbox\@tempboxa{{\captionfonts #1: #2}}%
 \ifdim \wd\@tempboxa >\hsize
    {\captionfonts #1: #2\par}
  \else
    \hbox to\hsize{\hfil\box\@tempboxa\hfil}%
  \fi
  \vskip\belowcaptionskip}
\makeatother   

\newcommand{\nn}{\nonumber}
\newcommand{\rom}[1]{\mathrm{#1}}

\newcommand\cA{\mathcal{A}}
\newcommand\cB{\mathcal{B}}
\newcommand\cC{\mathcal{C}}
\newcommand\cD{\mathcal{D}}
\newcommand\cE{\mathcal{E}}
\newcommand\cF{\mathcal{F}}
\newcommand\cG{\mathcal{G}}
\newcommand\cH{\mathcal{H}}
\newcommand\cI{\mathcal{I}}
\newcommand\cJ{\mathcal{J}}
\newcommand\cK{\mathcal{K}}
\newcommand\cL{\mathcal{L}}
\newcommand\cN{\mathcal{N}}
\newcommand\cO{\mathcal{O}}
\newcommand\cP{\mathcal{P}}
\newcommand\cQ{\mathcal{Q}}
\newcommand\cR{\mathcal{R}}
\newcommand\cS{\mathcal{S}}
\newcommand\cT{\mathcal{T}}
\newcommand\cU{\mathcal{U}}
\newcommand\cW{\mathcal{W}}
\newcommand\cX{\mathcal{X}}
\newcommand\cY{\mathcal{Y}}

\usepackage{mciteplus}

\usepackage[colorlinks=true,      linkcolor=blue,      urlcolor=blue,      filecolor=blue,      citecolor=blue,
      pdfstartview=FitH,     pdfpagemode=UseNone,      bookmarksopen=true]{hyperref}  

\usepackage[all]{hypcap}     

\begin{document}

\numberwithin{equation}{section}


\newcommand{\be}{\begin{equation}}
\newcommand{\ee}{\end{equation}}
\newcommand{\bea}{\begin{eqnarray}\displaystyle}
\newcommand{\eea}{\end{eqnarray}}

\def\eq#1{(\ref{#1})}


\def\a{\alpha}  \def\b{\beta}   \def\c{\chi}
\def\g{\gamma}  \def\G{\Gamma}  \def\e{\epsilon}
\def\vep{\varepsilon}   \def\tvep{\widetilde{\varepsilon}}
\def\f{\phi}    \def\F{\Phi}  \def\fb{{\ov \phi}}
\def\vf{\varphi}  \def\m{\mu}  \def\mub{\ov \mu}
\def\n{\nu}  \def\nub{\ov \nu}  \def\o{\omega}
\def\O{\Omega}  \def\r{\rho}  \def\k{\kappa}
\def\kab{\ov \kappa}  \def\s{\sigma}
\def\t{\tau}  \def\th{\theta}  \def\sb{\ov\sigma}  \def\S{\Sigma}
\def\l{\lambda}  \def\L{\Lambda}  \def\p{\psi}

\newcommand{\gt}{\tilde{\gamma}}


\def\cA{{\cal A}} \def\cB{{\cal B}} \def\cC{{\cal C}}
\def\cD{{\cal D}} \def\cE{{\cal E}} \def\cF{{\cal F}}
\def\cG{{\cal G}} \def\cH{{\cal H}} \def\cI{{\cal I}}
\def\cJ{{\cal J}} \def\cK{{\cal K}} \def\cL{{\cal L}}
\def\cM{{\cal M}} \def\cN{{\cal N}} \def\cO{{\cal O}}
\def\cP{{\cal P}} \def\cQ{{\cal Q}} \def\cR{{\cal R}}
\def\cS{{\cal S}} \def\cT{{\cal T}} \def\cU{{\cal U}}
\def\cV{{\cal V}} \def\cW{{\cal W}} \def\cX{{\cal X}}
\def\cY{{\cal Y}} \def\cZ{{\cal Z}}

\def\mC{\mathbb{C}} \def\mP{\mathbb{P}}
\def\mR{\mathbb{R}} \def\mZ{\mathbb{Z}}
\def\mT{\mathbb{T}} \def\mN{\mathbb{N}}
\def\mH{\mathbb{H}} \def\mX{\mathbb{X}}

\def\one{{\hbox{\kern+.5mm 1\kern-.8mm l}}}


\newcommand{\bra}[1]{{\langle {#1} |\,}}
\newcommand{\ket}[1]{{\,| {#1} \rangle}}
\newcommand{\braket}[2]{\ensuremath{\langle #1 | #2 \rangle}}
\def\corr#1{\left\langle \, #1 \, \right\rangle}
\def\vac{|0\rangle}


\def\d{ \partial }
\newcommand{\floor}[1]{\left\lfloor #1 \right\rfloor}
\newcommand{\ceil}[1]{\left\lceil #1 \right\rceil}

\mathchardef\mhyphen="2D

\newcommand{\T}[3]{\ensuremath{ #1{}^{#2}_{\phantom{#2} \! #3}}}		

\def\nn{\nonumber}

\def\sqi{{1\over \sqrt{2}}}
\newcommand\com[2]{[#1,\,#2]}


\newcommand{\orbifoldrefs}{Vafa:1995zh,Strominger:1996sh}

\newcommand{\orbifoldrecent}{Pakman:2009zz,*Pakman:2009ab,*Pakman:2009mi,*Keller:2011xi,*Cardona:2014gqa}

\newcommand{\twocharge}{Balasubramanian:2000rt,*Maldacena:2000dr,Lunin:2001jy,*Lunin:2002bj,Skenderis:2006ah,*Kanitscheider:2006zf,*Kanitscheider:2007wq}

\newcommand{\fuzzrecentone}{Bena:2011uw,*Giusto:2011fy,*Gibbons:2013tqa}

\newcommand{\fuzzrecenttwo}{Mathur:2013nja,*Bena:2013ora}

\newcommand{\fuzzrecentthree}{Bena:2014qxa,*Chen:2014loa,*Martinec:2014gka}

\newcommand{\lmone}{Lunin:2001ew}
\newcommand{\lmtwo}{Lunin:2001pw}
\newcommand{\acmtwo}{Avery:2010er}
\newcommand{\acmthree}{Avery:2010hs}


\begin{flushright}
IPHT-T15/144
\end{flushright}

\vspace{7mm}

\begin{center}
{\huge \textsc{Holographic description of} } \\

\vspace{6mm}

{\huge \textsc{non-supersymmetric orbifolded}}

\vspace{6mm}

{\huge \textsc{D1-D5-P solutions}}

\vspace{15mm}

{\large
\textsc{ Bidisha Chakrabarty${}^{1}$, ~ David Turton${}^{2}$, ~ Amitabh Virmani${}^{1, 3}$
}}

\vspace{13mm}

${}^{1}${Institute of Physics, Sachivalaya Marg, \\ Bhubaneshwar, India 751005}\\

\vspace{4mm}

${}^{2}${Institut de Physique Th{\'e}orique, CEA Saclay, CNRS URA 2306\\
91191 Gif-sur-Yvette, France}

\vspace{4mm}

${}^{3}${Max Planck Institute for Gravitational Physics (Albert Einstein Institute),\\ Am M\"uhlenberg 1,  D-14476 Potsdam-Golm,
Germany} \\

\vspace{8mm}

{\small\upshape\ttfamily bidisha@iopb.res.in, david.turton@cea.fr, virmani@iopb.res.in} \\

\vspace{18mm}

\textsc{Abstract}

\end{center}

\begin{adjustwidth}{10mm}{10mm} 

\vspace{0.3 cm}

{\small
\noindent
Non-supersymmetric black hole microstates are of great interest in the context of the black hole information paradox.
We identify the holographic description of the general class of non-supersymmetric orbifolded D1-D5-P supergravity solutions found by Jejjala, Madden, Ross and Titchener.
This class includes both completely smooth solutions and solutions with conical defects, 
and in the near-decoupling limit these solutions describe degrees of freedom in the cap region.
The CFT description involves a general class of states obtained by fractional spectral flow in both left-moving and right-moving sectors, generalizing previous work which studied special cases in this class.
We compute the massless scalar emission spectrum and emission rates in both gravity and CFT and find perfect agreement, thereby providing strong evidence for our proposed identification.  
We also investigate the physics of ergoregion emission as pair creation for these orbifolded solutions.
Our results represent the largest class of non-supersymmetric black hole microstate geometries 
with identified CFT duals presently known.
}

\end{adjustwidth}

\thispagestyle{empty}

\newpage

\baselineskip=13.5pt
\parskip=3pt

\tableofcontents

\baselineskip=15pt
\parskip=3pt

\section{Introduction}

The black hole information paradox~\cite{Hawking:1976ra} is a profound and long-standing problem in quantum gravity. String theory has had many successes in black hole physics, including the microscopic derivation of the entropy of the large supersymmetric D1-D5-P black hole~\cite{Strominger:1996sh}. The evidence from constructions of black hole microstates in string theory points to a resolution of the information paradox whereby the true quantum bound state has a size of order the event horizon of the naive classical solution, and so
the black hole event horizon and interior are replaced by quantum degrees of freedom. This is known as the fuzzball conjecture~\cite{Lunin:2001jy,Mathur:2005zp,*Bena:2007kg,*Skenderis:2008qn,*Balasubramanian:2008da,*Chowdhury:2010ct,Mathur:2012zp,*Bena:2013dka}.

To probe the quantum degrees of freedom of the black hole, one often studies semi-classical microstates, which may be described within supergravity.
There has been significant progress in the study of three-charge BPS black hole microstates~\cite{Giusto:2004id,*Lunin:2004uu,Giusto:2004ip,Bena:2004de,*Bena:2005va,*Berglund:2005vb,*Bena:2006kb,Giusto:2011fy,*Giusto:2012jx,*Giusto:2013rxa,Bena:2011uw,*Niehoff:2012wu,*Niehoff:2013kia}, culminating in the recent first explicit construction of a `superstratum'~\cite{Bena:2015bea}. 

Given a supergravity solution with the same charges as a black hole, it is important to establish whether the solution describes a bound state. In an AdS/CFT setup~\cite{Maldacena:1997re,Witten:1998qj,Gubser:1998bc}, one can do this by identifying a state in the holographically dual CFT. At present, there are many more such supergravity solutions than there are solutions with identified CFT duals. In the case of the superstratum, there is a proposal for the dual CFT states, evidence for which has recently been obtained~\cite{Giusto:2015dfa} using precision holography techniques~\cite{Skenderis:2006ah,*Kanitscheider:2006zf}.

Non-supersymmetric black hole microstate solutions are technically much more demanding; relatively few families have been explicitly constructed, and fewer still have known dual CFT states. The first non-supersymmetric black hole microstate solutions to be discovered were the solutions found by Jejjala, Madden, Ross and Titchener (often abbreviated to JMaRT)~\cite{Jejjala:2005yu}. For other studies of non-supersymmetric black hole microstate solutions, see~\cite{Gimon:2007ps,*Giusto:2007tt,*AlAlawi:2009qe, Bena:2011fc,*Bena:2012zi,Mathur:2013nja,Banerjee:2014hza,*Katsimpouri:2014ara,Bossard:2014yta,*Bossard:2014ola,Gibbons:2013tqa,*deLange:2015gca}.

The JMaRT family of solutions includes a positive integer parameter $k$; the solutions with $k>1$ can be thought of as orbifolds of the $k=1$ solutions, the orbifold acting on the asymptotic S$^1$ coordinate $y$. It is important to note that the $k=1$ solutions, while all smooth, do not exhaust the smooth solutions within this family, and that a significant parameter space of $k>1$ solutions are also smooth~\cite{Jejjala:2005yu}. In addition to the smooth solutions, there are also solutions with a rich possible structure of orbifold singularities; we will discuss this in detail in due course. 

In this paper we identify the CFT duals of the general class of orbifolded JMaRT solutions. Physically, the $k>1$ states are of particular interest; 
although the whole family of CFT states we study are atypical, states with larger $k$ are closer to typical states than states with smaller $k$.
This is because the typical three-charge state is in the maximally twisted sector, $k=n_1 n_5$, so states with higher $k$ are closer to typicality.

We work in the D1-D5 system on T$^4$. In the holographically dual orbifold CFT~\cite{\orbifoldrefs,Maldacena:1997re}, it has been proposed that
semiclassical states obtained by the action of the superconformal algebra generators on Ramond-Ramond (R-R) ground states are dual to bulk solutions involving diffeomorphisms that do not vanish at the boundary of the AdS throat~\cite{Mathur:2011gz,*Mathur:2012tj,*Lunin:2012gp}. Similarly, solutions involving non-trivial deformations (with respect to a reference R-R ground state) in the region deep inside the AdS throat known as the `cap' should be dual to CFT states that cannot be expressed in terms of superconformal algebra generators acting on R-R ground states; examples of three-charge BPS states which support this proposal were found in~\cite{Giusto:2012yz}. 

The CFT states studied in~\cite{Giusto:2012yz} involve fractional spectral flow in the left-moving sector, starting from the twisted R-R ground states studied in~\cite{Balasubramanian:2000rt,*Maldacena:2000dr}. The dual geometries are the BPS orbifold solutions found in~\cite{Giusto:2004ip,Jejjala:2005yu}.
It was anticipated in~\cite{Giusto:2012yz} that applying fractional spectral flow in both left- and right-moving sectors of the CFT, one should obtain states dual to the general orbifolded JMaRT geometries. In this paper we confirm this expectation, make precise the map between gravity and CFT, and provide strong evidence for the identification by studying the emission spectrum and emission rates of the states in both gravity and CFT. 
Although our main interest is in R-R states, the general class of CFT states we study also contains NS-NS states.

Since non-BPS, non-extremal states may be expected to be generically unstable, it is far from clear how many states might be described by stationary supergravity solutions. However the decay of such states is an opportunity to gain insight into the unitary mechanism that should replace Hawking radiation for generic states. 
In the case of the $k=1$ JMaRT solutions, soon after the discovery of these solutions it was shown that these geometries decay via a classical ergoregion instability~\cite{Cardoso:2005gj}.

A microscopic dual CFT explanation of the instability was proposed in~\cite{Chowdhury:2007jx}: the unitary CFT process of Hawking radiation is enhanced for the atypical CFT states dual to the JMaRT solutions, such that it manifests in the bulk as the ergoregion instability. Certain aspects of the CFT arguments were somewhat heuristic at the time, but were later made more precise in a series of papers \cite{Chowdhury:2008bd, Chowdhury:2008uj, Avery:2009tu, Avery:2009xr}. The spectrum and emission rate of minimal scalars from the microscopic considerations were found to be in exact agreement with the instability found on the gravity side. In this paper we extend these studies to the general $k>1$ case.

Finally, we explore the physical picture of ergoregion emission as pair creation \cite{friedman1978,*CominsSchutz}. 
This picture was investigated in reference \cite{Chowdhury:2008bd} for the two-charge $k=1$ JMaRT solutions. It
was shown that to a good approximation, radiation from these solutions
can be split into two distinct parts. One part escapes to infinity, and the other remains deep inside the AdS region, at the cap. In the present work, we generalize this picture to include all three charges
and the orbifolding parameter $k$, and consider the most general form of the probe scalar wavefunction.  
We confirm that also in this more elaborate set-up, the radiation splits into two distinct parts: one part escapes to infinity and the other part remains deep inside in the AdS region.

Our results generalize various previous studies (already mentioned above) of both BPS and non-BPS states arising from spectral flow of R-R ground states. We comment in detail on the relation of our work to these previous works after we have introduced the CFT states in full detail in Section \ref{sec:CFT_description}.

There has been a resurgence of interest in the black hole information paradox in recent years, in particular with regard to the experience of an infalling observer; see~\cite{Mathur:2009hf,Braunstein:2009my,Almheiri:2012rt,*Almheiri:2013hfa,Mathur:2012jk,Mathur:2013gua,Martinec:2014gka,Mathur:2015nra} and references within. Our results develop further the AdS/CFT dictionary for non-BPS black hole microstates, and such technical progress may ultimately shed light on these questions.

The remainder of this paper is organized as follows. In Section \ref{sec:JMaRT} we study the general family of orbifolded JMaRT solutions and solve the wave equation on these backgrounds.
In Section \ref{sec:CFT_description} we identify the CFT description of these geometries. The emission spectrum and rates obtained from the CFT are shown to be in perfect agreement with the gravity computation. In Section \ref{pair_creation} we analyze the pair creation picture of ergoregion emission for these orbifolds. We close with a brief discussion in Section \ref{sec:discussion}.

\section{Orbifolded JMaRT solutions}
\label{sec:JMaRT}

After a brief review of the supergravity solutions in Section \ref{sec:sugra}, we study the near-decoupling limit  in Section \ref{sec:largeR} in which the geometries have a large AdS inner region, weakly coupled to flat asymptotics. In Section \ref{sec:smoothness} we analyze the smoothness properties and categorize the possible orbifold singularities of the solutions. In Section \ref{sec:scalar_wave} we study the scalar wave equation on these orbifolds. We obtain the real and imaginary parts of the instability eigen-frequencies in the near-decoupling limit.

\subsection{Supergravity solutions}
\label{sec:sugra}

The JMaRT solutions \cite{Jejjala:2005yu} are special cases of the non-extremal rotating three-charge Cveti\v{c}-Youm \cite{Cvetic:1996xz} solutions.
In general, Cveti\v{c}-Youm geometries can have singularities, horizons, and closed timelike curves. Reference \cite{Jejjala:2005yu} derived the conditions that need to be imposed on the parameter space of the Cveti\v{c}-Youm geometries so that we get smooth solitonic solutions, possibly with orbifold singularities.

We consider type IIB string theory compactified on
\be
M_{4,1}\times \mathrm{S}^1\times {\mathrm T}^4 \,.
\label{compact}
\ee
We consider the S$^1$ to be macroscopic, and consider the T$^4$ to be string-scale. We consider $n_1$ D1-branes wrapped on S$^1$, $n_5$ D5-branes wrapped on S$^1\times \rom{T}^4$, and $n_p$ units of momentum P along the S$^1$. We parameterize the S$^1$ with coordinate $y$ and the T$^4$ with coordinates $z^i$.

Our supergravity analysis begins with the general non-extremal three-charge Cveti\v{c}-Youm metric, lifted to type IIB supergravity~\cite{Cvetic:1996xz,Cvetic:1997uw}. The 10D string frame metric is \cite{Jejjala:2005yu} 
\begin{eqnarray} \label{3charge}
ds^2&=&{}-\frac{f}{\sqrt{\tilde{H}_{1} \tilde{H}_{5}}}(
dt^2 - dy^2) +\frac{M}{\sqrt{\tilde{H}_{1}
\tilde{H}_{5}}} (s_p dy - c_p
dt)^2 \nonumber \\ 
&&{}+\sqrt{\tilde{H}_{1} \tilde{H}_{5}}
\left(\frac{ r^2 dr^2}{ (r^2+a_{1}^2)(r^2+a_2^2) - Mr^2}
+d\theta^2 \right)\nonumber \\ 
&&{}+\left( \sqrt{\tilde{H}_{1}
\tilde{H}_{5}} - (a_2^2-a_1^2) \frac{( \tilde{H}_{1} + \tilde{H}_{5}
-f) \cos^2\theta}{\sqrt{\tilde{H}_{1} \tilde{H}_{5}}} \right) \cos^2
\theta d \psi^2 \nonumber \\ 
&& {}+\left( \sqrt{\tilde{H}_{1}
\tilde{H}_{5}} + (a_2^2-a_1^2) \frac{(\tilde{H}_{1} + \tilde{H}_{5}
-f) \sin^2\theta}{\sqrt{\tilde{H}_{1} \tilde{H}_{5}}}\right) \sin^2
\theta d \phi^2 \nonumber \\ 
&&{} +
\frac{M}{\sqrt{\tilde{H}_{1} \tilde{H}_{5}}}(a_1 \cos^2 \theta
d \psi + a_2 \sin^2 \theta d \phi)^2 \nonumber \\
&&{}+ \frac{2M \cos^2 \theta}{\sqrt{\tilde{H}_{1} \tilde{H}_{5}}}[(a_1
c_1 c_5 c_p -a_2 s_1 s_5 s_p) dt + (a_2 s_1
s_5 c_p - a_1 c_1 c_5 s_p) dy ] d\psi \nonumber \\
&&{}+\frac{2M \sin^2 \theta}{\sqrt{\tilde{H}_{1} \tilde{H}_{5}}}[(a_2
c_1 c_5 c_p - a_1 s_1
s_5 s_p) dt + (a_1
s_1 s_5 c_p - a_2 c_1 c_5 s_p) dy] d\phi \cr
&&{} + \sqrt{\frac{\tilde{H}_{1} }{\tilde{H}_{5}}} \sum_{i=1}^{4}dz_i^2 \,,
\label{CYbig}
\end{eqnarray}
where we use the shorthand notation $c_i = \cosh \delta_i$, $s_i = \sinh \delta_i$, for $i =1,5,p$, and where
\begin{eqnarray} 
f~=~r^2+a_1^2\sin^2\theta+a_2^2\cos^2\theta\,,  \qquad
\tilde{H}_{1} ~=~  f+M\sinh^2\delta_1 \,, \qquad
\tilde{H}_{5} ~=~  f+M\sinh^2\delta_5 \,.
\eea
Explicit expressions for the six-dimensional dilaton and Ramond-Ramond two-form field can be found, e.g.,~in~\cite{Jejjala:2005yu}; we will not need those details in our discussion below. 
Upon compactification to five dimensions, one obtains asymptotically flat configurations carrying three U(1) charges, corresponding to D1, D5, and P. These charges are given by $Q_i = M s_i c_i.$

The $y$ circle will play a key role in the following; we take it to have 
radius $R$ at spacelike infinity, $y \sim y + 2 \pi R$.
In addition, we take the volume of T$^4$ to be $(2\pi)^4 V$ at spacelike infinity. The integer quantization of the three charges is then given by
\begin{align}
Q_1 &= \frac{g_s \alpha'{}^3}{V} n_1, &
Q_5 &= g_s \alpha' n_5, &
Q_p &= \frac{g_s^2 \alpha'{}^4}{V R^2} n_p. \label{integer_quantization}
\end{align}

The ADM mass and angular momenta of the five-dimensional asymptotically flat configurations are
\bea
M_\rom{ADM} &=&  \frac{\pi M }{4 G_5} \left( s_1^2 + s_5^2  + s_p^2  + \frac{3}{2} \right), \label{MassADM} \\
J_\psi &=& -    \frac{\pi M }{4 G_5} (a_1 c_1 c_5 c_p - a_2 s_1 s_5 s_p),  \label{jpsi} \\
J_\phi &=&  -   \frac{\pi M }{4 G_5} (a_2 c_1 c_5 c_p - a_1 s_1 s_5 s_p), \label{jphi}
\eea
where $G_5$ is the five-dimensional Newton constant. The ten-dimensional Newton constant is as usual $G_{10} = 8 \pi^6 g_s^2 \alpha'{}^{4}$, so we have $G_5 = \frac{\pi g_s^2 \alpha'{}^{4}}{4 V R}.$
To have positive ADM mass we take $M \ge 0$, and without loss of generality we take $\delta_1, \delta_5, \delta_p \ge 0$ and $a_1 \ge a_2 \ge 0$.

The singularities $H_1 = 0$ and $H_5=0$ in metric \eqref{CYbig} are curvature singularities, and there are also singularities where 
the function
\be
g(r) \equiv (r^2 + a_1^2)(r^2 + a_2^2) - M r^2 
\ee
has roots, i.e.~at
\be \label{eq:roots}
r_\pm^2 = \frac{1}{2}\left[(M - a_1^2 - a_2^2) \pm \sqrt{(M - a_1^2 - a_2^2)^2 - 4 a_1^2 a_2^2}\right].
\ee
Smooth geometries without horizons are obtained by demanding that at $r=r_+$ an $S^1$ should shrink smoothly, with the singularity at $r=r_+$ being that of polar coordinates at the origin of a two-dimensional factor of the metric~\cite{Jejjala:2005yu}. 
The parameter analysis is slightly different for the two-charge ($Q_p=0$) and three-charge cases; in this paper we focus on the general case of three non-vanishing charges. In this case four conditions on the parameters must be satisfied for the geometries to be smooth (up to possible orbifold singularities).  We now present a brief summary of the analysis of~\cite{Jejjala:2005yu}; for further details we refer the reader to that reference.

The function $g(r)$ has real roots if and only if $M > (a_1 + a_2)^2$ or $M < (a_1 - a_2)^2$. 
For $r=r_+$ to be an origin, rather than a horizon, the determinant of the metric in the constant $t$ and $r$ subspace must vanish at $r=r_+$. This rules out the case $M > (a_1 + a_2)^2$ and gives the first condition on the parameters,
\be
M= a_1^2 + a_2^2 - a_1 a_2 \frac{c_1^2 c_5^2 c_p^2 + s_1^2 s_5^2 s_p^2}{s_1 c_1 s_5 c_5 s_p c_p} \,.
\label{mass_condition}
\ee
In order that $r=r_+$ be an origin, a spacelike Killing vector with closed orbits must smoothly degenerate there. The most general Killing vector with closed orbits is 
\be
\xi_\rom{Killing} = \partial_y - \alpha \partial_\psi - \beta \partial_\phi,
\ee
and the one that degenerates at $r=r_+$,  given the condition \eqref{mass_condition}, is given by
\begin{align}
\alpha &= - \frac{s_p c_p }
{(a_1 c_1 c_5 c_p-a_2 s_1 s_5 s_p)}, & \beta &= - \frac{s_p c_p }
{(a_2 c_1 c_5 c_p- a_1 s_1 s_5 s_p)}.
\end{align}
We introduce new coordinates appropriate to the neighborhood of $r=r_+$,
\begin{align}
\bar \psi &\equiv \psi - \frac{s_p c_p }
{(a_1 c_1 c_5 c_p - a_2 s_1 s_5 s_p)} y,  &
\bar \phi \equiv \phi -
\frac{s_p c_p }
{(a_2 c_1 c_5 c_p - a_1 s_1 s_5 s_p)} y.
\label{eq:psiphibar}
\end{align}
Then the coordinate which shrinks at $r=r_+$ is $y$ at constant $\bar \psi$, $\bar \phi$.

In order to find the most general smooth solutions and to allow for the possibility of orbifold singularities at $r=r_+$, we introduce a positive integer $k$ and impose that $y \to y + 2 \pi k R$ at constant $\bar \psi$, $\bar \phi$, is a closed orbit.
This gives two further conditions,
\begin{align}
\frac{s_p c_p }
{(a_1 c_1 c_5 c_p- a_2 s_1 s_5 s_p)} (kR) &= n\in \mZ \,,  & -\frac{s_p c_p }
{(a_2 c_1 c_5 c_p- a_1 s_1 s_5 s_p)} (kR) &= m \in \mZ \,.
\end{align}
Furthermore, demanding  regularity at the origin $r=r_+$ under $y \to y + 2 \pi k R$  fixes the size of the $y$-circle at infinity,
\bea \label{eq:Rconstraint-k}
R &=& \frac{1}{k}\frac{M s_1 c_1 s_5 c_5 (s_1 c_1 s_5 c_5 s_p c_p)^{1/2}}
{\sqrt{a_1 a_2}(c_1^2 c_5^2 c_p^2 - s_1^2 s_5^2 s_p^2)}\,.
\eea

To summarize, the full regularity conditions for the three-charge orbifolded case are
\bea
&& (a) \qquad
a_1 a_2 ~=~ \frac{Q_1 Q_5}{k^2 R^2}\frac{s_1^2 c_1^2 s_5^2 c_5^2 s_p c_p} {(c_1^2 c_5^2 c_p^2 - s_1^2 s_5^2 s_p^2)^2},  \label{cond1} \\
&&
(b) \qquad
M~=~a_1^2 + a_2^2 - a_1 a_2 \frac{c_1^2 c_5^2 c_p^2 + s_1^2 s_5^2 s_p^2}{s_1 c_1 s_5 c_5 s_p c_p},  \label{cond2}  \\
&& (c) \qquad
\frac{s_pc_p}{(a_1 c_1 c_5 c_p - a_2 s_1 s_5 s_p)}(kR) ~=~n ~\in~\mathbb{Z}, \label{cond3}  \\
&& (d) \qquad
- \frac{s_pc_p}{(a_2 c_1 c_5 c_p - a_1 s_1 s_5 s_p)}(kR) ~=~ m ~\in~\mathbb{Z}.  \label{cond4}
\eea

Using these conditions we record here some relations between the parameters which will be useful in what follows,
\begin{align}
r_+^2 &=-a_1a_2 \frac{s_1 s_5 s_p}{c_1 c_5 c_p}, &
r_-^2 &=- a_1 a_2 \frac{c_1 c_5 c_p}{s_1 s_5 s_p}, &
M &= a_1 a_2 n m \left(\frac{c_1 c_5 c_p}{s_1 s_5 s_p} - \frac{s_1 s_5 s_p}{c_1 c_5 c_p}\right)^2.
\label{useful_eqs}
\end{align}
The ADM angular momenta, in terms of the parameters introduced above, take the following simple form,
\begin{align}
J_{\psi}  &= -\frac{m}{k} n_1n_5, &
J_{\phi} &= \frac{n}{k} n_1n_5.
\label{Jsimple}
\end{align}

\subsection{The near-decoupling limit }
\label{sec:largeR}

In order to study AdS/CFT in this system 
one must isolate low-energy excitations of the D1-D5 bound state, which is achieved by taking the large $R$ limit. In the gravity description, this corresponds to taking the  near-decoupling limit in which one obtains a large inner region involving an $\mathrm{AdS}_3 \times \mathrm{S}^3 \times \mathrm{T}^4$ throat, weakly coupled to the flat asymptotics.

The large $R$ limit is defined by keeping $Q_1, Q_5$ fixed and taking $ R \gg (Q_1 Q_5)^{\frac{1}{4}}$, which makes $R$ the largest scale in the problem. We then have the small dimensionless parameter
\bea
\epsilon = \frac{(Q_1 Q_5)^\frac{1}{4}}{R} ~  \ll ~ 1.
\label{epsilon}
\eea
In the Cvetic-Youm metric, the near-decoupling limit is obtained by taking
\bea
a_1^2, a_2^2, M \ll Q_1, Q_5  \qquad \Rightarrow \qquad s_1 \simeq c_1 \gg 1 \,, \quad  s_5 \simeq c_5 \gg 1. \label{approx}
\eea
We refer to the limit \eqref{epsilon}, \eq{approx} as the large $R$ limit or the near-decoupling limit.

In this limit we can identify the region
$
r^2 \ll Q_1, Q_5
$
as an asymptotically AdS region. This amounts to taking $\tilde H_1 \approx Q_1$ and $\tilde H_5 \approx Q_5$ and
using approximations \eqref{approx} in the metric \eqref{CYbig}. We thus obtain an asymptotically AdS$_3 \times \mathrm{S}^3$ metric,
\begin{eqnarray}
ds^2 &=& - \left(\rho^2 - M_3 +
\frac{J_3^2}{4 \rho^2} \right) d\tau^2 + \left(\rho^2
- M_3 + \frac{J_3^2}{4 \rho^2} \right)^{-1} d\rho^2 + \rho^2 \left(
d\varphi - \frac{J_3}{2\rho^2} d\tau \right)^2  \nn \\ &&
+ \sqrt{Q_1 Q_5}  \left\{ d\theta^2 
+ \sin^2 \theta \left[d\phi + \frac{R}{\sqrt{Q_1 Q_5}} (a_1 c_p - a_2
s_p) d\varphi + \frac{R}{\sqrt{Q_1 Q_5}} ( a_2
c_p-a_1 s_p) d \tau \right]^2
\right. \quad 
\nn \\
&&
{} \qquad\quad~~ {} \left.
+  \cos^2 \theta \left[d \psi + \frac{R}{\sqrt{Q_1 Q_5}}
(a_2 c_p - a_1 s_p) d\varphi + \frac{R}{\sqrt{Q_1 Q_5}}
(a_1 c_p-a_2 s_p) d \tau\right]^2 \right\} \label{AAdS}
\end{eqnarray}
where we have defined new coordinates
\begin{equation}
\varphi = \frac{y}{R}, \quad \tau = \frac{t}{R},
\label{varphi_tau_equation}
\end{equation}
\begin{equation}
\rho^2 = \frac{R^2}{Q_1 Q_5} [r^2 + (M-a_1^2-a_2^2) \sinh^2 \delta_p +
a_1 a_2 \sinh 2\delta_p ].
\label{rho_equation}
\end{equation}
The AdS length and the size of the S$^3$ is $(Q_1 Q_5)^{\frac{1}{4}}$. In writing the above expressions we have also defined
\begin{equation}
M_3 = \frac{R^2}{Q_1 Q_5} [ (M-a_1^2-a_2^2) \cosh 2\delta_p + 2 a_1 a_2
  \sinh 2 \delta_p],
\end{equation}
\begin{equation}
J_3 = \frac{R^2}{Q_1 Q_5} [ (M-a_1^2-a_2^2) \sinh 2\delta_p + 2 a_1 a_2
  \cosh 2 \delta_p].
\end{equation}

The regularity conditions \eqref{cond1}--\eqref{cond4} simplify in the large $R$ limit as follows,
\bea
&& (a') \qquad
a_1 a_2 ~\simeq~ \frac{Q_1 Q_5}{k^2 R^2} s_p c_p,  \label{scond1} \\
&&
(b') \qquad
M~\simeq~a_1^2 + a_2^2 - a_1 a_2 \frac{ c_p^2 +  s_p^2}{ s_p c_p},  \label{scond2}  \\
&& (c') \qquad
\frac{s_pc_p}{c_1 c_5 (a_1 c_p - a_2 s_p)}(kR) ~\simeq~n ~\in~\mathbb{Z}, \label{scond3}  \\
&& (d') \qquad
- \frac{s_pc_p}{c_1 c_5(a_2  c_p - a_1  s_p)}(kR) ~\simeq~ m ~\in~\mathbb{Z}.  \label{scond4}
\eea
A useful form of condition \eqref{scond2} via \eqref{useful_eqs} is
\be
M \simeq \frac{Q_1 Q_5}{(kR)^2} \frac{n m}{s_p c_p},
\ee
and another expression that will be useful later is
\be
r_+^2 - r_-^2 \simeq \frac{Q_1 Q_5}{k^2 R^2}.
\label{rplusrminus}
\ee
Substituting conditions \eqref{scond1}--\eqref{scond4} into \eqref{AAdS} we find that the geometry is an orbifold of AdS$_3 \times \mathrm{S}^3$,
\bea
ds^2 &=& \sqrt{Q_1 Q_5} \left[ - \left(\frac{1}{k^2}+ \rho^2\right) d\tau^2 + d\rho^2\left(\frac{1}{k^2}+\rho^2\right)^{-1} + \rho^2 d \varphi^2 \right. \nn \\
&&
\label{metric_decoupled}
\left. + d\theta^2 + \sin^2 \theta \left(d\phi + \frac{m}{k} d\varphi - \frac{n}{k} d\tau \right)^2
+ \cos^2 \theta \left(d\psi - \frac{n}{k} d\varphi + \frac{m}{k} d\tau\right)^2 \right].
\eea
We will analyze the smoothness properties of these orbifold geometries in the next subsection.

Let us now look at how various physical quantities behave in the large $R$ limit. It is only in this limit that we expect physical parameters in the gravity description to be reproduced by a dual CFT analysis.
In this limit the mass above the mass of the D1 and D5 branes is
\be
\Delta M_\rom{ADM}  \simeq
\frac{\pi M}{4 G_5} \left( s_p^2 + \frac12 \right)
\simeq  \frac{n_1 n_5}{R} \frac{m^2 + n^2 - 1}{2 k^2},
\label{Msimple}
\ee
and the 6D ADM linear momentum $P_\rom{ADM}$ is
\be
P_\rom{ADM} = \frac{n_p}{R}  =\frac{\pi M}{4 G_5} s_p c_p
\simeq  \frac{n_1 n_5}{R}\frac{m n}{k^2}.
\label{PADM}
\ee
In Section \ref{sec:CFT_description} we will observe agreement between the gravity quantities \eqref{Jsimple}, \eqref{Msimple},  and \eqref{PADM} from the CFT description.

\subsection{Smoothness analysis}
\label{sec:smoothness}

As observed above, the decoupling limit of the general JMaRT orbifolded solution is an orbifold of AdS$_3 \times$ S$^3$, with metric \eqref{metric_decoupled}.
These orbifolds were briefly discussed in \cite{Jejjala:2005yu}; here we present a detailed smoothness analysis following \cite{Giusto:2012yz}.
For convenience let us introduce the coordinates
\begin{align}
\tilde \psi &\equiv \psi - \frac{n}{k} \varphi + \frac{m}{k} \tau, &
\tilde \phi &\equiv \phi + \frac{m}{k} \varphi - \frac{n}{k} \tau. &
\end{align}
Working in the covering space for coordinates $(\varphi,  \tilde \psi, \tilde \phi)$,
the periodicities of $(\varphi,  \psi, \phi)$ translate into the following identifications:
\begin{eqnarray}
A: && (\varphi, \tilde \psi, \tilde \phi) \to  (\varphi, \tilde \psi, \tilde \phi) + 2 \pi \left(1, -\frac{n}{k},  \frac{m}{k} \right),\\
B: && (\varphi, \tilde \psi, \tilde \phi) \to  (\varphi, \tilde \psi, \tilde \phi) + 2 \pi (0,1,0),\\
C: && (\varphi, \tilde \psi, \tilde \phi) \to  (\varphi, \tilde \psi, \tilde \phi) + 2 \pi (0,0,1).
\end{eqnarray}

Note that the coordinate $\varphi$ in the metric \eqref{metric_decoupled} is ill-defined at $\rho=0$, where a conical singularity
can occur; the periodicity required for smoothness is $\varphi \to \varphi + 2\pi k$ at fixed $\tilde \psi, \tilde \phi$. Conical singularities only occur at points that remain invariant under the operation
\be
A^{m_A} B^{m_B} C^{m_C}\label{generators}
\ee
for some $m_I \in \mathbb{Z}$. 
The conical singularities all arise at $\rho=0$ and may be localized at $\theta=0$ and/or $\theta=\frac{\pi}{2}$, or may occur everywhere in $\theta$.
We will continue to focus on the case of three non-zero charges; the two-charge case is discussed in \cite{Jejjala:2005yu}.


\subsubsection*{\underline{Case 1:  $\gcd(k,m)=\gcd(k,n)=1$}}
If there are no common divisors between the pairs $(m,k)$ and $(n,k)$, then there are no conical singularities and the spacetime is completely smooth.
To see this, we first examine the possibility of having a fixed point 
where $\tilde \phi$ has a non-zero size, i.e.,~at $\rho=0, \theta \neq 0$. For a fixed point to occur here, $\tilde{\phi}$ must remain invariant under \eqref{generators}. This implies that $\frac{m}{k} m_A + m_C = 0$. Since $m_C$ is an integer this requires $\frac{m_A}{k}$  to be an integer; we write $m_A=km_A^{\prime}$. 
The periodic identifications of $\varphi $ and $\tilde{\psi}$  are then
\bea
\varphi &\rightarrow  &\varphi + 2 \pi km_A^{\prime} \,, \qquad\quad 
\tilde{\psi} ~\rightarrow ~  \tilde{\psi} + 2 \pi \left(m_B- n m_A^{\prime}\right)  . \label{psi_diffeo}
\eea
In the range $0 < \theta < \frac{\pi}{2}$,  $\tilde \psi$ also has a finite size. So for a fixed point to occur there, $\tilde \psi$  must also remain invariant.
This fixes $m_B=n m_A^{\prime}$, and as a result the above identification becomes
\bea
\varphi &\rightarrow  &\varphi + 2 \pi k m_A^{\prime} \,, \label{varphi_diffeo}
\eea
which, being an integer multiple of $2 \pi k$, is the correct identification for smoothness.

At $\theta = \frac{\pi}{2}$, $\tilde \psi$ has zero size. Thus, under the diffeomorphism \eqref{psi_diffeo} the point $\rho = 0, \theta = \frac{\pi}{2}$ is a fixed point. 
The relevant identification is simply \eq{varphi_diffeo} and we again have smoothness.

It remains to examine  $\rho=0 , \theta=0$.
Here $\tilde \phi$ has zero size but $\tilde \psi$ has non-zero size. Requiring $\tilde \psi$ to be invariant fixes $- \frac{n}{k} m_A + m_B = 0$. Since $m_B$ is an integer, this implies $m_A$ should be an integer multiple of $k$; we again write it as $m_A = k m_A^{\prime}$.
The relevant identification is again \eqref{varphi_diffeo}. This shows that the spacetime is free of conical singularity here also.

In summary, there are no conical singularities 
anywhere, and so the spacetime is completely smooth. From the point of view of the $k=1$ JMaRT solitons, one can say that the $\mathbb{Z}_k$ quotient is freely acting in this case \cite{Jejjala:2005yu}.

\subsubsection*{\underline{Case 2:  $ \gcd(k,m)>1, \  \gcd(k,n) =1$}}
If $ \gcd(k,m)\equiv l_1>1$ and $\gcd(k,n) =1$, there is a $\mathbb{Z}_{l_1}$ orbifold singularity at $\rho=0, \theta=\frac{\pi}{2}$ and the spacetime is otherwise smooth.

To see this, we first note that at $\rho =0, \theta = 0$, the analysis is the same as in Case 1, and there is no orbifold singularity at these points.

Next, let us write $k=l_1 \hat{k}\,$, $m=l_1 \hat{m}$. For a fixed point at $\rho=0, \theta \neq 0$, $\tilde{\phi}$ must remain invariant. This fixes $m_A = \hat{k} m_A^{\prime}\,$, $m_C = - \hat{m} m_A^{\prime} $. At points $\theta \neq \frac{\pi}{2}$, $\tilde{\psi}$ also has a non-zero size, so it must also remain invariant. This fixes $- \frac{n}{k} (\hat k m_A^{\prime}) + m_B = 0$, i.e, $m_B = \frac{n}{l_1} m_A ^{\prime}$. Since $m_B$ is an integer, $m_A ^{\prime}$ must be an integer multiple of $l_1$, i.e., $m_A ^{\prime} = l_1 m_A ^{\prime \prime}$. Then the $\varphi$ identification
$\varphi \to \varphi + 2 \pi m_A$ becomes $\varphi \to \varphi + (2 \pi k) m_A^{\prime \prime}$ since we have
\be
m_A = \hat k m_A^{\prime}
=  \hat k l_1 m_A^{\prime \prime}
=  k m_A^{\prime \prime}.
\ee
Hence we have smoothness at $\rho=0, \, 0<\theta< \frac{\pi}{2}$.

For $\rho=0, \theta = \frac{\pi}{2}$, $\tilde \psi$ has zero size. Invariance of $\tilde{\phi}$ gives $m_A = \hat{k} m_A^{\prime}$. So the $\varphi$ identification
$\varphi \to \varphi + 2 \pi m_A$ becomes $
\varphi \to \varphi + 2 \pi (\hat k m_A^{\prime})$, i.e.,
\bea
\varphi &\rightarrow& \varphi + (2 \pi k) \frac{m_A^{\prime}}{l_1}.
\eea
Since ${m_A^{\prime}}$ is a general integer, there is a $\mathbb{Z}_{l_1}$ orbifold singularity at $\rho=0, \theta=\frac{\pi}{2}$.

\subsubsection*{\underline{Case 3:  $ \gcd(k,m)= 1 ,  \ \gcd(k,n)  >1$}}

If $\gcd(k,m) =1$ and $\gcd(k,n)\equiv l_2>1$, there is a $\mathbb{Z}_{l_2}$ orbifold singularity at $\rho=0, \theta=0$ and the spacetime is otherwise smooth.
To see this, firstly an analysis similar to Case 2 shows that there are no conical singularities at  $\rho=0, \theta = \frac{\pi}{2}$ or at $\rho=0, \, 0< \theta < \frac{\pi}{2}$.

Next, let us write $k =  l_2 \hat{k} $, $n=l_2 \hat{n}$. For $\rho=0, \theta= 0$ to be a fixed point $\tilde{\psi}$ must remain invariant.  This fixes $m_A=\hat{k}m_A^{\prime }, m_B=\hat{n} m_A^{\prime}$. 
The $\varphi$ identification $\varphi \to \varphi+ 2\pi m_A$ then becomes $\varphi \to \varphi+ 2 \pi \hat{k} m_A^{\prime }$, which is
\bea
\varphi &\rightarrow& \varphi+ (2 \pi k) \frac{m_A^{\prime}}{l_2}.
\eea
This results in a $\mathbb{Z}_{l_2}$ orbifold singularity at $\rho=0, \theta=0$.

\subsubsection*{\underline{Case 4:  $ \gcd(k, m)>1,  \  \gcd(k, n)  >1, \ \gcd(k,m,n) = 1$}}
When both $\gcd(k,m)\equiv l_1>1$ and $\gcd(k,n)\equiv l_2>1$, the spacetime has both a $\mathbb{Z}_{l_1}$ orbifold singularity at $\rho=0, \theta=\frac{\pi}{2}$ and a  $\mathbb{Z}_{l_2}$ orbifold singularity at $\rho=0, \theta=0$. Away from these points the metric is smooth.
The analysis is similar to the previous two cases.

\subsubsection*{\underline{Case 5:  $ \gcd(k, m)>1,  \  \gcd(k, n)  >1 \ \gcd(k,m,n) > 1$}}
When $\gcd(k,m)\equiv l_1>1$, $\gcd(k,n)\equiv l_2>1$, and $\gcd(k,m,n) \equiv l_3 > 1$, then the orbifold has a rich singularity structure with
\begin{itemize}
\item $\mathbb{Z}_{l_1}$ orbifold singularity at $\rho=0, \, \theta=\frac{\pi}{2}$,
\item $\mathbb{Z}_{l_2}$ orbifold singularity at $\rho=0, \, \theta=0$, and
\item $\mathbb{Z}_{l_3}$ orbifold singularity at $\rho=0, \, 0< \theta <  \frac{\pi}{2}$.
\end{itemize}
Thus at $\rho=0$ there is at least a $\mathbb{Z}_{l_3}$ orbifold singularity 
all over the three-sphere\footnote{Note that the orbifold singularity at $\rho=0, \, 0< \theta <  \frac{\pi}{2}$ only arises in the non-BPS case, since in the BPS limit we have $m=n+1$ and so $m$ and $n$ have no common divisors.},
which may be enhanced to a singularity of higher degree at the poles if $l_1$ and/or $l_2$ are greater than $l_3$.

To see this, we first observe that an analysis similar to Cases 2 and 3  shows that there is a $\mathbb{Z}_{l_1}$ orbifold singularity at $\rho=0, \, \theta=\frac{\pi}{2}$ and a $\mathbb{Z}_{l_2}$ orbifold singularity at $\rho=0, \, \theta=0$. 

To see the orbifold singularity at $\rho=0, \, 0< \theta <  \frac{\pi}{2}$ we introduce the following notation,
\begin{align}
m &= l_1 \hat m, &
n &= l_2 \hat n, &
l_1 &= l_3 \hat l_1, &
l_2 &= l_3 \hat l_2, &
k &= l_3 \hat l_1 \hat l_2 \hat k. &
\end{align}
To have a fixed point at $\rho=0, \, 0< \theta <  \frac{\pi}{2}$, both $\tilde \psi$ and $\tilde \phi$  must remain invariant. From the invariance of $\tilde \phi$ we get
\bea
\frac{m}{k}m_A + m_C &=& 0  \qquad \Rightarrow \qquad \frac{\hat m}{\hat l_2 \hat k}m_A + m_C ~=~ 0 \,.
\eea
Since $m_C$ is an integer, $m_A$ must be a multiple of $\hat l_2 \hat k$, so we write $m_A = \hat l_2 \hat k m_A^\prime.$ This gives $m_C=-\hat m m_A^\prime.$

Similarly, from  the invariance of $\tilde \psi$ we get
\bea
-\frac{n}{k}m_A + m_B &=& 0  \qquad \Rightarrow \qquad -\frac{\hat n \hat l_2}{\hat l_1}m_A^\prime + m_B ~=~ 0.
\eea
Since $m_B$ is an integer, $m_A^\prime$ must be a multiple of $\hat l_1$, i.e., $ m_A^\prime = \hat l_1  m_A^{\prime \prime}.$
This implies $ m_A = \hat l_1 \hat l_2 \hat k m_A^{\prime \prime}, $ and $m_B = \hat n \hat l_2 m_A^{\prime \prime}.$
The $\varphi$ identification now becomes
\bea
\varphi &\rightarrow& \varphi + 2 \pi \hat l_1 \hat l_2 \hat k m_A^{\prime \prime} = \varphi + (2 \pi k) \frac{m_A^{\prime \prime}}{l_3}.
\eea
Hence there is a $\mathbb{Z}_{l_3}$ orbifold at $\rho = 0$ and $\theta \neq 0, \theta \neq \frac{\pi}{2}$.

We finally note that although we presented the above analysis for the decoupled asymptotically AdS$_3 \times \mathrm{S}^3$ geometries, it applies equally well to the asymptotically flat geometries before taking the decoupling limit using \eq{eq:psiphibar}.

\subsection{Scalar wave equation}
\label{sec:scalar_wave}

We next study a minimally coupled scalar in six dimensions on the general orbifolded JMaRT solutions. For the $k=1$ solutions, such a computation showed  that these geometries suffer from a classical ergoregion instability~\cite{Cardoso:2005gj}.
We extend this study to the case of general $k,m,n$, obtaining the real and imaginary parts of the instability eigen-frequencies in the large $R$ limit. We will later reproduce these results from the CFT.

Let us consider a minimally coupled complex scalar $\Psi$ in six dimensions, on the background of the dimensionally reduced 6D Einstein frame metric. 
If one takes the 10D string frame metric written in \eq{CYbig} and discards the $\mathrm{T}^4$ directions, one obtains exactly the 6D Einstein frame metric, and so we will not rewrite it here.
Such a minimal scalar arises for example
from the dimensional reduction of the ten-dimensional IIB graviton having both its indices along the four-torus. We can separate variables using the ansatz,
\be
\Psi = \exp\left[- i \omega t + i m_\psi \psi + i m_\phi \phi + i \frac{\lambda}{R}y\right] \chi (\theta) h(r),
\label{Psi_main_text}
\ee
which gives equation for the angular part
\be
\frac{1}{\sin 2\theta} \frac{d}{d\theta} \left( \sin 2 \theta \frac{d}{d\theta} \chi \right) + \left[ \left( \omega^2 - \frac{\lambda^2}{R^2}\right) (a_1^2 \sin^2 \theta + a_2^2 \cos^2 \theta)
- \frac{m_\psi^2}{\cos^2 \theta} - \frac{m_\phi^2}{\sin^2\theta}\right] \chi = -\Lambda \chi. \label{angular}
\ee
We are looking for wave functions with frequency $\omega \sim \frac{1}{R}$. In the large $R$ limit,
in terms of $\epsilon$ defined in \eq{epsilon}, we observe that
\bea
\left( \omega^2 - \frac{\lambda^2}{R^2}\right) a_i^2 \;\sim\; \epsilon^4
\eea
and so we find
\bea
\Lambda &=& l (l+2) + \mathcal{O}(\epsilon^4) \,.
\eea
The radial equation takes the form
\bea
& &\frac{1}{r} \frac{d}{dr} \left( \frac{g(r)}{r} \frac{d}{dr} h \right) - \Lambda h +
\left[
\left(\omega^2 - \frac{\lambda^2}{R^2}\right) (r^2 + M s_1^2 + M s_5^2)  +
\left(\omega c_p - \frac{\lambda}{R} s_p\right)^2 M
\right] h \nn \\
 & & - k^2 \frac{r_+^2 - r_-^2}{r^2 - r_+^2} \left(-\lambda - \frac{n}{k} m_\psi + \frac{m}{k} m_\phi \right)^2 h + k^2 \frac{r_+^2 - r_-^2}{r^2 - r_-^2} \left(\omega \varrho R - \lambda \vartheta - \frac{n}{k} m_\phi + \frac{m}{k} m_\psi \right)^2 h  = 0, \nn \\
\eea
where $g(r) = (r^2 - r_+^2)(r^2 - r_-^2)$. Introducing the dimensionless variable $x$ for the radial coordinate via
\be \label{eq:x-def}
x = \frac{1}{k^2} \left( \frac{r^2 - r_+^2}{r_+^2 - r_-^2} \right) ,
\ee
we can write the radial equation in the form
\be
\partial_x \left[ x\left(x+\frac{1}{k^2}\right) \partial_x h \right] + \frac{1}{4}\left[\kappa^2 x + 1 - \nu^2 + \frac{\xi^2}{x+k^{-2}} - \frac{\zeta^2}{x}\right]h = 0,
\label{radial_eq}
\ee
with
\bea
\kappa^2 &=& \left(\omega^2 - \frac{\lambda^2}{R^2}\right)(r_+^2 -r_-^2) k^2, \label{kappa}\\
\xi &=& \omega \varrho R - \lambda \vartheta - m_\phi \frac{n}{k} + m_\psi \frac{m}{k}, \label{xi} \\
\zeta &=& - \lambda - m_\psi \frac{n}{k} + m_\phi \frac{m}{k}, \label{zeta} \\
\varrho &=& \frac{c_1^2 c_5^2 c_p^2  - s_1^2 s_5^2 s_p^2}{s_1 c_1 s_5 c_5}, \label{varrho} \\
\vartheta &=& \frac{c_1^2 c_5^2  - s_1^2 s_5^2}{s_1 c_1 s_5 c_5} s_p c_p, \label{vartheta} \\
\nu^2 &=& 1 + \Lambda - \left(\omega^2 - \frac{\lambda^2}{R^2}\right)  (r_+^2 + M s_1^2 + M s_5^2) - \left(\omega c_p - \frac{\lambda}{R}s_p\right)^2 M.
\label{nu}
\eea
For later use, we note from \eqref{nu} that the correction to $\nu$ is $\mathcal{O}(\epsilon^2)$,
\bea
\nu &=& l+1 + \cO(\epsilon^2)\,.
\label{nu_large_R}
\eea
The radial differential equation \eqref{radial_eq} cannot be solved exactly. It can however be solved via matched asymptotic expansion.
This is done in detail in Appendix \ref{app:wave}. The instability frequencies are given by solutions to the transcendental equation \eqref{matching}. To the leading order in the large $R$ expansion, we let one of the $\Gamma$ functions in the denominator of \eqref{matching} develop a pole,
\begin{equation}
\frac{1}{2}(1+\nu + k |\zeta| + k \xi) \simeq -N,
\label{pole_main}
\end{equation}
with $N$ a non-negative integer.   From equations $\eqref{varrho}$ and \eqref{vartheta} we see that
in the large $R$ limit $\varrho \to 1$ and $\vartheta \sim \epsilon^2$.
Hence to leading order one obtains
\bea
\xi &\simeq& \omega R - m_\phi \frac{n}{k} + m_\psi \frac{m}{k}\,.  \label{xilargeR}
\eea
Substituting this relation along with \eq{nu_large_R} into equation \eqref{pole_main}, we get the real part $\omega_R$ of the instability frequencies to leading order, which are given by
\bea
\omega_R &\simeq&
\frac{1}{k R } \Bigl( - l - m_\psi m  + m_\phi n  -  \left|-  k \lambda - m_\psi n + m_\phi m \right| -   2(N + 1) \Bigr).
\label{omegaRFinal_maintext}
\eea
For certain values of the parameters, $\omega_R$ can become negative or zero. For those cases there is no emission.

One obtains the imaginary part $\omega_I$ of the instability frequencies to leading order by iterating the above approximation to the next order, setting $N \to N + \delta N$. This computation is discussed in detail in Appendix \ref{app:wave}. The result is
\bea
\omega_I &\simeq&
\frac{1} {kR} \frac{\pi}{2^{2 l + 1}(l!)^2}
\left[\left(\omega^2-\frac{\lambda^2}{R^2}\right) \frac{Q_1 Q_5}{k^2 R^2} \right]^{l+1}
{N + l + 1\choose l+1} {N+k| \zeta | + l + 1 \choose l+1} \,.  \qquad
\label{omegaIFinal_maintext}
\eea
Since $\omega_I > 0$, we have an instability: i.e., an exponentially growing perturbation. In the following section we reproduce \eqref{omegaRFinal_maintext} and \eqref{omegaIFinal_maintext} 
from the dual CFT.

\section{CFT description of orbifolded JMaRT solutions}
\label{sec:CFT_description}

\subsection{The D1-D5 system on T$^4$ and the orbifold CFT}

In order to discuss the CFT interpretation of the general orbifolded JMaRT solutions, we next review some properties of the D1-D5 system on $\rom{T}^4$ and the corresponding orbifold CFT. We follow in places the presentations of~\cite{Carson:2014ena} and \cite{Giusto:2012yz}.

As mentioned in the previous section, we work in type IIB string theory compactified on $M_{4,1}\times \rom{S}^1\times \rom{T}^4 $, with $n_1$ D1-branes wrapped on $\rom{S}^1$ and $n_5$ D5-branes wrapped on $\rom{S}^1\times \rom{T}^4$.
We work in the limit of large $R$, which corresponds to the low-energy limit of the gauge theory on the D-brane bound state.

At low energies, the gauge theory on the bound state flows to a $(4,4)$ SCFT. It is conjectured that there is a point in moduli space where this SCFT is a symmetric product orbifold theory, consisting of $n_1n_5$ symmetrized copies of a free $(4,4)$ SCFT with target space $\rom{T}^4$ \cite{\orbifoldrefs}.

Each copy of $\rom{T}^4$ gives 4 bosonic fields $X^1, X^2, X^3, X^4$, along with 4 left-moving fermionic excitations $\psi^1, \psi^2, \psi^3, \psi^4$ and the corresponding right-moving excitations, which we denote with a bar ($\bar{\psi}^1$, etc.).  The total central charge of the CFT is $c=6 n_1n_5$.

The CFT has a (small) $\mathcal{N}=4$ superconformal symmetry in both the left and right-moving sectors. Each superconformal algebra contains an R-symmetry $SU(2)$.  Therefore we have the global symmetry $SU(2)_L\times SU(2)_R$, whose quantum numbers we denote as
\be\label{ExternalQuantumNumbers}
SU(2)_L: ~(j_L, m_L);~~~~~~~SU(2)_R: ~ (j_R, m_R).
\ee
In addition there is a broken $SO(4) \simeq SU(2)\times SU(2)$ symmetry,
corresponding to rotations in the four directions of the $\rom{T}^4$.  We label this symmetry by
\be
SU(2)_1\times SU(2)_2.
\ee
We use indices $\alpha, \dot\alpha$ for $SU(2)_L$ and $SU(2)_R$ respectively, and indices $A, \dot A$ for $SU(2)_1$ and $SU(2)_2$ respectively.  The 4 real fermion fields of the left sector are grouped into complex fermions $\psi^{\alpha A}$. The right fermions are grouped into fermions $\bar{\psi}^{\dot\alpha A}$. The boson fields $X^i$ are a vector in $\rom{T}^4$ and have no charge under $SU(2)_L$ or $SU(2)_R$, so are grouped as $X_{A \dot A}$. Different copies of the $c=6$ CFT are denoted with a copy label in brackets, e.g.,~
\bea
X^{(1)} \,, ~ X^{(2)} \,,~ \cdots \,, ~X^{(n_1n_5)} \,.
\eea

It will be convenient to describe the states of interest in terms of spectral flow~\cite{Schwimmer:1986mf}.
Under a spectral flow transformation on the left-moving sector with parameter $\alpha$, the dimensions and charges of states change as follows:
\be
h'=h+\alpha m_L +\alpha^2 \frac{c}{24} \,, \qquad\quad
m_L'=m_L+\alpha \frac{c}{12} \,.
\label{eq:spectral}
\ee
An independent spectral flow operation exists in the right-moving sector, with parameter $\bar \alpha$.

\subsection{Twisted Ramond sector ground states}

We next briefly review the construction of twist operators and twisted Ramond sector ground states by mapping to a local covering space \cite{\lmtwo,Carson:2014ena}.
Let us consider the permutation $(123\ldots k)$.
The bare twist operator $\s_k$ corresponding to this permutation imposes the following periodicity conditions on the cylinder:
\bea
X^{(1)} \rightarrow X^{(2)} \rightarrow \cdots \rightarrow X^{(k)} \rightarrow X^{(1)} ~~ \label{eq:bp1} \cr
\psi^{(1)} \rightarrow \psi^{(2)} \rightarrow \cdots \rightarrow \psi^{(k)} \rightarrow \; -\psi^{(1)}. \label{eq:fp1}
\eea
Note that the last sign in the second line above is minus, and is the only physically meaningful sign, as the intermediate signs can be absorbed by field redefinitions\footnote{One of the authors (DT) thanks Oleg Lunin for a discussion on this point.}. This state is then in the NS sector in the covering space, which we will sometimes refer to simply as the NS sector. A similar expression holds for the right-moving fermions; for ease of presentation we will write only the left-moving expressions in various places in the following. It is convenient to describe these $k$ twisted copies of the CFT as a `component string' of length $k$.

One defines the bare twist operator $\s_k$ by mapping 
first to the plane with coordinate $z=e^w$
and then to a local covering plane with coordinate $t$ via a map of the local form
\bea
z-z_* \approx b_* \left(t-t_* \right)^k,
\label{eq:locmap}
\eea
where $z_*$ and $t_*$ are the respective images of $w_*$ in the $z$ plane and the $t$ plane.  
The $k$ bosonic fields in \eq{eq:bp1} map to one single-valued bosonic field $X(t)$ in the $t$ plane, and similarly for the fermions.
In the $t$ plane, one inserts the identity operator at the point $t_*$, obtaining the lowest-dimension operator in the $k$-twisted sector.
If we take $t_*=0$, we obtain the NS-NS vacuum in the covering space. 
We thus refer to it as the ``$k$-twisted NS-NS vacuum'', and denote it by $\ket{0_k}_{\rom{NS}}^{(r)}$, where $r$ is an index labelling the different component strings. The quantum numbers of this state are 
\be
h ~=~ \bar h ~=~ \frac{1}{4}\left( k - \frac{1}{k} \right) \,, \qquad\quad m_L ~=~ m_R ~=~ 0 \,.
\ee

We next define an excited (spin-)twist operator $\s^{\a}_k$ as follows.  Follow the procedure used to define the bare twist $\s_k$, but in the covering $t$ plane, insert a spin field\footnote{If $b_* \neq 1 $ in \eq{eq:locmap}, one must also include an appropriate normalization factor~\cite{\lmtwo}.} $S^{\a}$ at $t_*$. If we take $t_*=0$, we obtain the (left-moving)
R vacuum $|0_{\rom{R}}^{\pm} \rangle _t$ of the $t$ plane. We write
\bea
\s_k^{\a} = S_k^{\a} \sigma_k \,,   \qquad \a = +,- \,.
\eea
Back on the original cylinder, with coordinate $w$, as the fields circle the operator $\s_k^{\pm}$, they transform as
\bea
X^{(1)} \rightarrow X^{(2)} \rightarrow \cdots \rightarrow X^{(k)} \rightarrow X^{(1)}\cr
\psi^{(1)} \rightarrow \psi^{(2)} \rightarrow \cdots \rightarrow \psi^{(k)} \rightarrow +\psi^{(1)}.
\eea
The fields are thus in the Ramond sector in the covering space; as before, we will sometimes refer to this simply as the Ramond sector.
We write the corresponding state on the original cylinder (with coordinate $w$) 
as $|0_k^\pm\rangle_{\rom{R}}^{(r)}$.

Adding in the right-moving sector, we obtain the full spin-twist field
\bea
\s_k^{\alpha \dot \alpha} = S_k^{\alpha} \bar{S}_k^{\dot \alpha} \s_k
\eea
and we denote the corresponding twisted R-R ground state by $|0_k^{\alpha \dot \alpha}\rangle_{\rom{R}}^{(r)}$.

\subsection{Non-BPS states generated by general fractional spectral flow}

We now consider spectral flow operations in the $k$-fold covering space. 
Spectral flow by $\a_c$ units in the $k$-fold covering space corresponds to an effective spectral flow in the base space by an amount~\cite{Giusto:2012yz}
\bea
\a &=& \frac{\a_c}{k} \,.
\eea
On the base space, this may then be described as `fractional spectral flow'; for previous discussions of fractional spectral flow, see \cite{Martinec:2001cf,Martinec:2002xq,Avery:2009xr}.

Using this operation we now describe the general AdS/CFT dictionary for the $k>1$ JMaRT solutions. 
All of the states we consider consist of $n_c = N_1 N_5/k$ component strings of length $k$, with each component string in the same state\footnote{In this paper we consider parameters such that $n_c$ is an integer.}; spectral flow acts simultaneously on all component strings.
On a component string of length $k$, excitations are spaced in units of $1/k$. 
Fractional spectral flow generates states with filled Fermi seas with this fractional moding, as we will see explicitly shortly.

Let us first define the reference state from which we will perform the fractional spectral flows. This state has all of its component strings in the $k$-twisted NS-NS vacuum:
\bea\label{eq:kvac}
|0_k\rangle_{\rom{NS}} &=& \ket{0_k}_{\rom{NS}}^{(1)}\otimes \ket{0_k}_{\rom{NS}}^{(2)}
\otimes \cdots \otimes \ket{0_k}_{\rom{NS}}^{(n_c)} \,.
\eea
 The quantum numbers of this state are (here $c=6n_1n_5$ for the full CFT)
\be
h ~=~ \bar h ~=~  \frac{c}{24}\left( 1 - \frac{1}{k^2} \right) \,, \qquad\quad m_L ~=~ m_R ~=~ 0 \,.
\ee
The AdS dual of this state is the decoupled orbifolded AdS solution \eq{metric_decoupled} with $m=n=0$. The full asymptotically flat JMaRT solitonic solutions exist only when $|m| \neq |n|$ (if one works with $a_1 \ge a_2 \ge 0$, this becomes $m>n\ge 0$), so this solution does not directly come from the decoupling limit of an asymptotically flat JMaRT solution\footnote{It is however related to the other decoupled JMaRT solutions by (fractional) spectral flow coordinate transformations, which do not go to zero at the boundary of AdS.}.

The states we are interested in are obtained by general fractional spectral flow from $|0_k\rangle_{\rom{NS}}$. The map to the JMaRT solutions is that the spectral flow parameters are given by  
\bea
\a &=& \frac{m+n}{k} \,, \qquad \qquad \bar\a ~=~ \frac{m-n}{k} \,.
\label{parameter_map}
\eea
Using \eq{eq:spectral}, the quantum numbers of the spectral flowed states are
\begin{eqnarray}
h &=& \frac{c}{24}\left[1 + \frac{(m+n)^2 -1}{k^2} \right] , \qquad m_L ~=~ \frac{c}{12} \frac{m+n}{k} \,, \\
\bar h &=&  \frac{c}{24}\left[1 + \frac{(m-n)^2 -1}{k^2} \right] , \qquad
m_R ~=~ \frac{c}{12} \frac{m-n}{k} \,.
\end{eqnarray}
Therefore the CFT energy above the R-R ground state and momentum are
\bea
\Delta E ~=~ \frac{\Delta h+\Delta\bar h}{R} &=& \frac{n_1 n_5}{R} \frac{m^2 + n^2 - 1}{2 k^2}\,, 
\qquad\quad 
P ~=~ \frac{h-\bar h}{R}  ~=~ \frac{n_1 n_5}{R}\frac{m n}{k^2} \,.
\label{EP_CFT}
\eea
Note that in the orbifold CFT, the momentum on each component string must be an integer (see e.g.~\cite{Dijkgraaf:1996xw}), so in the orbifold CFT one has $mn/k\in\mZ$.

Using the map between CFT and gravity SU(2) quantum numbers,
\bea
m_\psi &=& -(m_L+ m_R)  \,, \qquad
 m_\phi ~=~  (m_L- m_R) \,,
\label{eq:cftSU2s}
\eea
these parameters exactly match those computed on the gravity side in \eq{Jsimple}, \eq{Msimple} and \eq{PADM}, providing a first check on our proposed identification. 

The above states are R-R in the covering space when $m+n$ is odd, and NS-NS in the covering space when $m+n$ is even.
Our main interest is in the R-R states; in order to connect with the discussion in~\cite{Giusto:2012yz}, let us present the free fermion description of these states, focussing on the states with positive $m_L$ and $m_R$.

Let us first consider a single component string. Recall that on a component string of length $k$, excitations are spaced in units of $1/k$. 
The state on the component string involves Fermi seas filled to a general fractional level $s/k$ in both species of fermions, $\psi^{+1}$ and $\psi^{+2}$, and similarly to a level $\bar s/k$ for the right-movers:
\bea\label{eq:FrSpFlSt-comp}
|\Phi_{s,\bar s,k}\rangle^{(r)}
&=& \left[
\left(\psi^{+1}_{-{ s\over k}}\psi^{+2}_{-{s\over k}}\right)\dots
   \left(\psi^{+1}_{-{2\over k}}\psi^{+2}_{-{2\over k}}\right)
	 \left(\psi^{+1}_{-{1\over k}}\psi^{+2}_{-{1\over k}}\right)
	\right] \cr
&& \qquad\qquad \times	\left[
\left(\bar\psi^{+1}_{-{ \bar s\over k}}\bar\psi^{+2}_{-{\bar s\over k}}\right)\dots
   \left(\bar\psi^{+1}_{-{2\over k}}\bar\psi^{+2}_{-{2\over k}}\right)
	 \left(\bar\psi^{+1}_{-{1\over k}}\bar\psi^{+2}_{-{1\over k}}\right)
	\right]
	|0^{++}_k\rangle_{\rom{R}}^{(r)} \,.
\eea
Then as before the state of the full CFT is obtained by taking all $n_c = N_1 N_5/k$ component strings to be in the same state:
\bea\label{eq:FrSpFlSt}
|\Psi_{s,\bar s,k}\rangle &=&  |\Phi_{s,\bar s,k}\rangle^{(1)} \otimes |\Phi_{s,\bar s,k}\rangle^{(2)}
 \otimes \cdots \otimes |\Phi_{s,\bar s,k}\rangle^{(n_c)} \,.
\eea

The twisted R-R ground state $|0^{++}_k\rangle_{\rom{R}}^{(r)}$ may be obtained from the twisted NS-NS vacuum $|0_k\rangle_{\rom{NS}}$ by performing fractional spectral flow with parameters $\a=1/k$, $\bar\a =1/k$.
The above state $|\Psi_{s,\bar s,k}\rangle$ is generated by a further fractional spectral flow with parameters $\a=2s/k$, $\bar\a =2s/k$. So in total, $|\Psi_{s,\bar s,k}\rangle$ is generated by starting with the state $|0_k\rangle_{\rom{NS}}$ and performing fractional spectral flow with parameters
\bea
\a &=& \frac{2s+1}{k} \,, \qquad\quad \bar\a ~=~ \frac{2\bar s+1 }{k}\,.
\eea
We then have the relations
\bea
m+n &=& 2s+1 \,, \qquad\qquad  m-n ~=~  2\bar s+1 \,.
\label{parameter_map-2}
\eea
The NS-NS states obtained for even $m+n$ have analogous Fermi sea representations, built on the twisted NS-NS vacuum $|0_k\rangle_{\rom{NS}}$.

We now return to our main discussion. For general $k,m,n$, we have observed the agreement of conserved charges above.
As was noted in \cite{Giusto:2012yz} in the BPS case however, generically these states are degenerate and so further evidence is required to support the identification. 
In principle, one could compute the one-point functions of operators following~\cite{Skenderis:2006ah,*Kanitscheider:2006zf}, however the states we are considering are R-charge eigenstates, and therefore all one-point functions of R-charged operators vanish~\cite{Skenderis:2006ah}. Instead,
we provide further evidence for our proposed identification by matching the scalar excitation spectrum between gravity and CFT.

\subsection{Emission spectrum and emission rates from CFT}

The vertex operator for emission (or absorption) of a minimal scalar of angular momentum $l$ has the following form~\cite{Avery:2009tu}. It involves a chiral primary in the twisted sector of degree $(l+1)$, $\tilde\sigma_{l+1}$, dressed with fermion and supercurrent excitations $G_{-{\frac12}}^{+\dot A}\psi_{-{\frac12}}^{-A}\bar{G}_{-{\frac12}}^{\dot{+}\dot B}\bar\psi_{\dot{-}{\frac12}}^{-B}$ which add the $\rom{T}^4$ polarization indices, and further dressed with powers of SU(2) current zero modes $J_0$, $\bar{J}_0$ which fill out the SU(2) representation. There is also a non-trivial normalization factor; the explicit form can be found in~\cite{Avery:2009tu}.

Since the vertex operator involves a twisted chiral primary $\tilde\sigma_{l+1}$, when it acts on a state it introduces new fractionated degrees of freedom. It is thus capable of lowering the energy of the state, with the remainder energy being carried away by the emitted particle.

Our initial state \eq{eq:FrSpFlSt-comp} is composed of component strings which are all of length $k$.
In the limit of a large number of component strings, $n_c=n_1n_5/k \gg 1$, the process which dominates is that in which $\tilde \sigma_{l+1}$ acts on $l+1$ distinct component strings, combining them into a component string of length $k(l+1)$. 
There is a family of resulting final states labelled by left- and right-moving excitation numbers $N_L$, $N_R$, which correspond to acting with the Virasoro generators $L_{-1}$, $\bar{L}_{-1}$ in the form $L_{-1}^{N_L} \bar{L}_{-1}^{N_R}$ on the final state of lowest possible energy.

This CFT amplitude, corresponding to emission of a minimal scalar, can be mapped to a technically simpler amplitude by spectral flow and hermitian conjugation. This technique has been employed in the special case of excited R-R states arising from integer spectral flow of the state $|0^{++}_k\rangle_{\rom{R}}$, i.e.,~when $s$ and $\bar s$ are multiples of $k$. This was first done for $k=1$ in~\cite{Avery:2009tu} and then for $k>1$ in~\cite{Avery:2009xr}. For the $k>1$ case, the calculation involved mapping the amplitude to a covering space, and using the method of~\cite{Lunin:2000yv,Lunin:2001pw}\footnote{For a recent application of this method in a different context, see~\cite{Carson:2014ena,Carson:2014yxa}.}. 
In each case one observes a Bose enhancement effect: the probability for emission of the $\mbox{\footnotesize{{\bf N}}}^{\rom{th}}$ quantum is $\mbox{\footnotesize{{\bf N}}}$ times the probability for emission of the first quantum~\cite{Chowdhury:2007jx}.
Since the CFT is a symmetric product orbifold, one must also take care of various combinatorial factors in computing the amplitude.

%
Having proposed the identification of the general orbifolded JMaRT solutions with the general fractional spectral flowed CFT states, we can now make a straightforward generalization of the results of \cite{Avery:2009xr} to fractional spectral flowed CFT states. We do this by simply taking the emission spectrum, expressed in terms of $\a$, $\bar\a$, and substituting the values appropriate for the general fractional spectral flowed states that we study. This technique works because all of the states under consideration are fractional spectral flows of 
the twisted NS-NS vacuum $|0_k\rangle_{\rom{NS}}$.

The emission spectrum computed in~\cite{Avery:2009xr} for the integer spectral flowed $k>1$ JMaRT states, translated into our conventions\footnote{The map between conventions is given in Appendix \ref{app:conventions}.}, is
\bea
\o &=& \frac{1}{kR} \left[
 \frac12 \a k ( m_\phi - m_\psi) - \frac12\bar\a k ( m_\phi + m_\psi)
-(l +2 + N_L+N_R)
\right], \cr
\l &=& \frac{1}{kR} \left[
 \frac12 \a k ( m_\phi - m_\psi) + \frac12\bar\a k ( m_\phi + m_\psi)
+ N_R-N_L
\right].
\label{eq:cftspectrum1}
\eea
We now generalize this by substituting the parameters appropriate to fractional spectral flow 
from the twisted NS vacuum $|0_k\rangle_{\rom{NS}}$,
\bea
\a &=& \frac{m+n}{k} \,, \qquad \qquad \bar\a ~=~ \frac{m-n}{k} \,,
\eea
which yields the spectrum
\bea
\o &=& \frac{1}{kR} \left[-m_\psi m+m_\phi n -(l +2 + N_L+N_R)\right] \cr
\l &=& \frac{1}{k} \left[ m_\phi m -  m_\psi n + N_R- N_L \right].
\label{eq:CFT-omega-lambda}
\eea
Now, generalizing the discussion in~\cite{Chowdhury:2007jx}, note that $\zeta=(N_L-N_R)/k$. If $\zeta>0$, $\o$ may be written as
\bea
\o &=& \frac{1}{kR} \left[-l -m_\psi m + m_\phi n - 2 - k \zeta - 2 N_R \right]
\eea
and if $\zeta<0$, $\o$ may be written as
\bea
\o &=& \frac{1}{kR} \left[-l -m_\psi m + m_\phi n - 2 + k \zeta - 2 N_L \right] .
\eea
In either case, $\o$ has the form
\be
\omega = \frac{1}{k R } \left( - l - m_\psi m  + m_\phi n  -  \left|-  k \lambda - m_\psi n + m_\phi m \right| -   2(N + 1) \right)
\label{omegaRFinal_maintext-1}
\ee
for some $N\ge 0$, which exactly matches the real part of the instability frequencies computed from the gravity side, given in Eq.\;\eq{omegaRFinal_maintext}.

The CFT emission rate computed in \cite{Avery:2009xr} for the
$\mbox{\footnotesize{{\bf N}}}^{\rom{th}}$
particle, writing $\delta(\omega;\lambda)$ as a schematic delta function which imposes that $\omega$ and $\lambda$ must take their specific allowed values, in our conventions takes the form
\bea
\frac{d\Gamma}{d\omega} &=& \mbox{\footnotesize{{\bf N}}} \frac{1}{kR}\frac{2\pi}{2^{2l+1}(l!)^2} \left[\left(\omega^2-\frac{\lambda^2}{R^2}\right) \frac{Q_1 Q_5}{k^2 R^2} \right]^{l+1}
{N_L + l + 1\choose l+1} {N_R + l + 1 \choose l+1} \delta(\omega;\lambda) \,.  \qquad
\eea
In this form, the expression for the emission rate immediately generalizes to the present situation of fractional spectral flowed states, with the allowed frequencies and wavelengths given in \eq{eq:CFT-omega-lambda}.

Treating separately the cases for $\zeta>0$ and $\zeta<0$ as above, one finds
\bea
{N_L + l + 1\choose l+1} {N_R + l + 1 \choose l+1} &=& {N + l + 1\choose l+1} {N+k| \zeta | + l + 1 \choose l+1} \,.
\eea

The imaginary part of the frequency $\omega_I$ is given by 1/2 the value of the emission rate for the first quantum, as discussed in \cite{Chowdhury:2007jx}. Thus we have
\bea
\omega_I &\simeq&
\frac{1} {kR} \frac{\pi}{2^{2 l + 1}(l!)^2}
\left[\left(\omega^2-\frac{\lambda^2}{R^2}\right) \frac{Q_1 Q_5}{k^2 R^2} \right]^{l+1}
{N + l + 1\choose l+1} {N+k| \zeta | + l + 1 \choose l+1} \,,  \qquad
\label{omegaIFinal_maintext-2}
\eea
in exact agreement with the value \eq{omegaIFinal_maintext} obtained from the gravity calculation.

\subsection*{Relation to previous work}

We pause here to comment on the relation of our results to previous literature.

The class of states generated by \eq{parameter_map} is the general set of R-R and NS-NS states obtained by fractional spectral flow from the twisted NS-NS vacuum $|0_k\rangle_{\rom{NS}}$. Various special cases of this class of CFT states have been studied previously in the literature, as we now describe.

For BPS states, the two-charge states $(k \in \mZ^+,m=1,n=0)$ were studied in~\cite{Balasubramanian:2000rt,*Maldacena:2000dr}. The three-charge family $(k=1; \, s \in \mZ; \, \bar s = 0 )$ was studied in~\cite{Giusto:2004id,*Lunin:2004uu}. The family $(k \in \mZ^+ ; \, s = n k, n \in \mZ ; \, \bar s = 0 )$ was studied in~\cite{Giusto:2004ip}. Such values of $s$ 
correspond to integer spectral flow of the states $|0^{\a\dot\a}_k\rangle_{\rom{R}}$.
The general BPS family obtained from fractional spectral flow, $(k \in \mZ^+ \; s\in \mZ ; \, \bar s=0 )$ was studied in~\cite{Giusto:2012yz}.

For non-BPS states, the CFT states obtained by setting $k=1$ in \eq{eq:kvac}--\eq{parameter_map} were proposed to be the dual CFT states of the $k=1$ JMaRT solutions in the original paper~\cite{Jejjala:2005yu} and the CFT emission was studied in \cite{Chowdhury:2007jx,Avery:2009tu}.
The two-charge family $(k \in \mZ^+; \, s = \bar s = \hat n k, \hat n \in \mZ )$ was studied in~\cite{Chowdhury:2008uj}.
The family $(k \in \mZ^+ ;\, s = \hat n k, \hat n\in \mZ; \, \bar s = \bar n k , \bar n \in \mZ )$ was studied in~\cite{Avery:2009xr}.
Again, such values of $s, \bar s$ correspond to integer spectral flow of the states $|0^{\a\dot\a}_k\rangle_{\rom{R}}$.
The general non-BPS family of R-R and NS-NS states arising from fractional spectral flow of $|0_k\rangle_{\rom{NS}}$ (or $|0^{\a\dot\a}_k\rangle_{\rom{R}}$) is the subject of the present work.

Regarding the wave equation calculation on the gravity side,
for $k=1$ the instability was first derived in \cite{Cardoso:2005gj}, and was revisited in slightly different forms in \cite{Chowdhury:2007jx,Chowdhury:2008bd}. The two-charge case with $k>1$ was studied in~\cite{Chowdhury:2008uj}. In the present work we have analyzed the general three-charge case with arbitrary $k,m,n$.

\section{Ergoregion emission as pair creation}
\label{pair_creation}

Having demonstrated that the general class of orbifolded JMaRT solutions decay via an ergoregion instability, with emission spectrum and emission rate in agreement with the dual CFT, we now examine more explicitly some features of the produced radiation. In particular we investigate the physical picture of ergoregion emission as pair creation~\cite{friedman1978,*CominsSchutz}. 

The ergoregion contains negative energy excitations as measured by the Killing vector that generates time translations at spatial infinity. The pair creation picture involves a positive energy excitation that escapes to infinity and a negative energy excitation that remains in the ergoregion. The two excitations also carry equal and opposite values of other conserved charges. 

For two-charge, $k=1$ JMaRT solutions, this picture was investigated in~\cite{Chowdhury:2008bd} for the simplest form of the probe scalar wavefunction.
It was shown that to a good approximation, the radiation from these solutions can be split into two distinct parts. One part escapes to infinity and the other remains deep inside in the AdS region. The two parts carry equal and opposite energy and angular momentum. For large angular momenta, when the wavefunctions can be thought of as approximately localized, it was argued that the inner region part has its main support in the ergoregion. 

In this section we generalize this discussion to include three non-zero charges, two non-zero angular momenta, the orbifolding parameter $k$, and the most general form of the wavefunction. We start with a summary of the solutions of the scalar wave equation in Section \ref{sec:solutions_wave_summary}. We then compute the contributions to angular momenta (Section \ref{sec:angular_momenta}) and energy (Section \ref{sec:energy}) from the inner and asymptotic regions due to the scalar perturbation.

\subsection{Solutions of the wave equation}
\label{sec:solutions_wave_summary}
In order to calculate the contributions to conserved charges from the inner and asymptotic regions (which are defined in Appendix \ref{app:wave}), we need the explicit form of the wavefunctions in these regions. We work exclusively in the large $R$ limit, as only in this limit is there a clear separation between the inner and asymptotic regions.

Let us start by relating the different radial coordinates so that we can easily change from one to the other. The coordinate transformation \eqref{rho_equation} upon using
\eqref{useful_eqs} and \eqref{scond2} is simply
\be
\rho^2 = \frac{R^2}{Q_1 Q_5} (r^2 - r_+^2) \,. \label{rho_inner}
\ee
In terms of the dimensionless radial variable $x$ used in Section \ref{sec:scalar_wave}, this relation is
$x = \rho^2.$

The metric in the inner region is \eqref{metric_decoupled}, and from \eqref{hinner} the wavefunction in the inner region is
\be
\Psi_\rom{in} =
 \exp\left[- i \omega t + i \frac{\lambda}{R}y + i m_\psi \psi + i m_\phi \phi \right]
\chi(\theta) \left(\rho^2 + \frac{1}{k^2}\right)^{\frac{k\xi}{2}} \rho^{k|\zeta|} \left[{}_2F_1(a,b,c,- k^2 \rho^2)\right], \label{hinner_main_text}
\ee
where
\bea
a &=& \frac{1}{2} (1+ \nu + k|\zeta| + k \xi), \qquad
b ~=~ \frac{1}{2}(1 -\nu + k |\zeta| + k \xi), \qquad
c ~=~ 1+ k |\zeta|,
\eea
with  $\xi$, $\zeta$, $\nu$ defined in \eqref{xi},  \eqref{zeta}, \eqref{nu}.
Recall also from \eqref{nu_large_R}, \eqref{pole_main} that to leading order in $\epsilon$ we have
\bea
\nu &\simeq& l + 1, \qquad\qquad
1+ \nu+ k |\zeta| + k \xi  ~\simeq~ - 2N. \label{integer_replacements}
\eea

For small $\rho$, we have $\Psi_\rom{in} \sim \rho^{k|\zeta|}$, and for large $\rho$, using \eq{inner_expansion} and \eq{integer_replacements} we have $\Psi_\rom{in}~\sim~\rho^{-  (l+2)}$.
The norm of the wavefunction is
\begin{equation}
(\Psi \Psi^*)_\rom{in} = \rho^{2k|\zeta|} \left(\rho^2 + \frac{1}{k^2}\right)^{k\xi}  |\chi(\theta)|^2
e^{2\omega_I t} \Bigl({}_2F_1(-N,-N-l-1,1+k |\zeta|,-k^2\rho^2)\Bigr)^2,
\label{Psi_norm}
\end{equation}
where as before $\omega_I$ is the imaginary part of the frequency $\omega$.

In the asymptotic region the metric is flat spacetime to leading order. Using the asymptotic region wavefunction \eqref{outer_wavefunction} together with the requirement of only outgoing waves \eqref{outgoing}, in terms of a  normalization constant $C_2$ and the quantity $\kappa$ defined in \eqref{kappa}, the wavefunction is
\be
\Psi_\rom{out} = C_2
\exp\left[- i \omega t + i m_\psi \psi + i m_\phi \phi + i \frac{\lambda}{R}y\right]
\chi(\theta)   \frac{1}{\sqrt{2\pi \kappa}} \frac{1}{\rho^{\frac{3}{2}}}e^{i \kappa \rho} e^{-i \frac{\pi}{4}}  \left(e^{i \frac{\pi \nu}{2}} - e^{-i \frac{3\pi \nu}{2}}\right) .
\label{houter_main_text}
\ee
Therefore the norm of the wavefunction is
\be
(\Psi \Psi^*)_\rom{out}
= |C_2|^2  |\chi(\theta)|^2  e^{2\omega_I t}
\frac{2}{\pi |\kappa|} \frac{1}{\rho^3} e^{i (\kappa -\kappa^*) \rho}  \sin^2 (\pi \nu) \,.
\label{hhstar_main_text}
\ee
We next fix the normalization of $\Psi$ in the asymptotic region given its form in the inner region.This is done in detail in Appendix \ref{app:pair_details}. For our purposes we do not need an expression for $C_2$ itself, but only its norm. From \eqref{normc2} we have
\be
|C_2|^2 \sin^2(\pi\nu) = \frac{ \pi}{2} k^{4 N + 2 \nu}(k R)\omega_I \,
\frac{\Gamma(1+k|\zeta|)^2  \  \Gamma(N+1) \ \Gamma(N+\nu+1)}{\Gamma(N+\nu +1+k|\zeta|) \ \Gamma(N+1+k|\zeta|)} \,,
\label{normc2_main_text}
\ee
where $\omega_I$ takes the value given in \eqref{omegaIFinal_maintext}.
Finally we note that in the asymptotic region
$
\rho^2~\simeq~\frac{R^2}{Q_1 Q_5} r^2,
$
and as a result the exponent in \eqref{hhstar_main_text} can be written as
\be
i (\kappa -\kappa^*) \rho \simeq -\frac{2\omega_R \omega_I}{\sqrt{\omega_R^2 - \frac{\lambda^2}{R^2}}} r \,,
\label{power}
\ee
where as before $\omega_R$ is the real part of the frequency $\omega$ and takes the value given in~\eq{omegaRFinal_maintext}.
In the neck region, the exponent \eqref{power} is very small,
$| (\kappa -\kappa^*) \rho|  \sim  \omega_I (Q_1 Q_5)^{1/4} \sim \epsilon^{4l+5}$.

\subsection{Angular momenta of the perturbation}
\label{sec:angular_momenta}

The general JMaRT solution has four Killing vectors, namely $\partial_t, \partial_y, \partial_\phi, \partial_\psi$. In general the geometries carry angular momentum in both $\phi$ and $\psi$
directions and momentum in the $y$ direction.
As in the previous sections, we consider scalar perturbations that also carry all these charges. The conserved quantities for the scalar perturbation associated to the two
angular momenta are
\bea
L_\psi &=& \int T_\psi{}^\nu dS_\nu, \qquad\qquad  
L_\phi ~=~ \int T_\psi{}^\nu dS_\nu, \label{angular_momentum}
\eea
where $T_\mu{}^\nu$ is the energy momentum tensor of the (complex) scalar field,
\be
T_{\mu \nu} = \partial_{\mu} \Psi \partial_{\nu} \Psi^* + \partial_{\nu} \Psi \partial_{\mu} \Psi^* - g_{\mu \nu} \partial_\alpha \Psi \partial^\alpha \Psi^*. \label{stress_tensor}
\ee
The integrals in \eqref{angular_momentum} extend over a spacelike hypersurface in the spacetime. We choose the surface to be simply given by $t=$ constant. It was shown in reference \cite{Jejjala:2005yu} that $g^{tt} < 0$ everywhere, therefore  the $t=$ constant surface is everywhere spacelike.

Substituting the separation ansatz \eqref{Psi_main_text} in \eqref{stress_tensor}, we find the following expressions for angular momenta of the scalar perturbation,
\begin{eqnarray}
 L_\psi &=&  2 m_\psi \int \sqrt{-g} dr d\mathcal{A} \left( -g^{tt} \omega_R + g^{t\psi} m_\psi + g^{t\phi} m_\phi + g^{ty} \frac{\lambda}{R} \right) \Psi \Psi^*,
\label{L_psi_simp}
\\
 L_\phi &=&  2 m_\phi \int \sqrt{-g} dr d\mathcal{A} \left( - g^{tt} \omega_R + g^{t\psi} m_\psi + g^{t\phi} m_\phi  + g^{ty} \frac{\lambda}{R} \right) \Psi \Psi^*,
\end{eqnarray}
where $d \cA = d \theta d\psi d\phi dy$. Note that the integrals involved in computing $L_\psi$ and $L_\phi$ are the same. For this reason we focus on $L_\psi$; the discussion for $L_\phi$ is entirely analogous.

In the asymptotic region the metric is flat spacetime to leading order. We have $\sqrt{-g} = r^3 \cos\theta \sin\theta$. There are no cross terms in the metric, so the integral \eqref{L_psi_simp} simply becomes
\bea
(L_\psi)_\rom{out} &=&  2 m_\psi \omega_R \int_\rom{out} dr d\mathcal{A}  (r^3 \cos\theta \sin\theta)  (\Psi \Psi^*)_{\rom{out}},
 \nn \\
&=& 4 \pi m_\psi \omega_R  R C e^{2\omega_I t} \int_\rom{out} dr  (r^3 h(r) h(r)^*)_\rom{out},\label{Lpsiout}
\eea
where 
$
C = \int_{\rom{S}^3} d\theta d\phi d\psi  \cos \theta \sin \theta |\chi(\theta)|^2.
$
Using relations \eqref{hhstar_main_text} and \eqref{power}  expression \eqref{Lpsiout} becomes
\bea
(L_\psi)_\rom{out} &=&\frac{8Q_1 Q_5 m_\psi \omega_R   C }{R \sqrt{\omega_R^2 - \frac{\lambda^2}{R^2}}} e^{2\omega_I t}  |C_2|^2 \sin^2 (\pi \nu)\int_{(Q_1 Q_5)^{\frac{1}{4}}}^{\infty} dr \exp\left(-\frac{2\omega_R \omega_I}{\sqrt{\omega_R^2 - \frac{\lambda^2}{R^2}}} r\right).
\eea
To leading order in the large $R$ limit this integral gives
\be
(L_\psi)_\rom{out} \simeq  2 \pi m_\psi Q_1 Q_5 C  e^{2\omega_I t}  k^{ 4 N + 2 l + 3}\frac{\Gamma(1+k|\zeta|)^2 \ \Gamma(N+1) \ \Gamma(N+l+2)}{\Gamma(N+k |\zeta|+l +2) \ \Gamma(N+k |\zeta| +1)},
\label{Lpsifinal}
\ee
where we have used the normalization \eqref{normc2_main_text}. This is our final expression for the angular momentum $L_\psi$  of the scalar perturbation that flows off to infinity.
For $N=0$, $k=1$, and $|\zeta| = 0$ this expression reduces to the corresponding expression of reference \cite{Chowdhury:2008bd}.

Exactly the same expression but with opposite sign is obtained from the inner region. Using the coordinate definitions \eqref{rho_inner} and \eqref{varphi_tau_equation}, and the metric in the inner region \eqref{metric_decoupled}, the integral
\eqref{L_psi_simp} in the inner region becomes
\be
(L_{\psi})_{\rom{in}} = 2 m_\psi  Q_1 Q_5 \left(\omega_R R + \frac{m}{k} m_\psi -\frac{ n}{k} m_\phi   \right)
\int\limits_{0}^{1/\epsilon} \rho d \rho \int d \cA \cos\theta\sin\theta \left(\rho^2 + \frac{1}{k^2}\right)^{-1}
( \Psi \Psi^*)_\rom{in}. \nn
\ee
Substituting the norm \eqref{Psi_norm} of the inner region wavefunction, we observe that the integrand falls off in the large $\rho$ limit as $\rho^{-2l -5}$. Thus to leading order in $\epsilon$ we can set the upper limit of the $\rho$ integration to infinity. Thus we obtain
\bea
(L_{\psi})_{\rom{in}}&\simeq& 4 \pi m_\psi Q_1 Q_5 C  e^{2\omega_I t}
\left(\omega_R R + \frac{m}{k} m_\psi -\frac{ n}{k} m_\phi   \right)
 \times \nn \\
 & &  \int_{0}^{\infty}  d \rho \rho^{2k|\zeta|+1} \left(\rho^2 + \frac{1}{k^2}\right)^{k \xi-1}
\Bigl({}_2F_1(-N,-N-l-1,1+k |\zeta|,-k^2\rho^2)\Bigr)^2.  \nn
\eea
Making the substitution $\tilde \rho = k \rho$ and using the integer relations \eqref{integer_replacements}, this expression can be converted to  the form
\bea
(L_{\psi})_{\rom{in}} &\simeq& 4 \pi m_\psi Q_1 Q_5 C  e^{2\omega_I t}
\left(\omega_R R + \frac{m}{k} m_\psi -\frac{ n}{k} m_\phi   \right) k^{ 4 N + 2 l + 4}
 \times \nn \\
 & &  \int_{0}^{\infty}  d \tilde \rho \tilde \rho^{2k|\zeta|+1} \left(\tilde \rho^2 + 1\right)^{- 2 N- l - k |\zeta| - 3}
\Bigl({}_2F_1(-N,-N-l-1,1+k |\zeta|,-\tilde \rho^2)\Bigr)^2. \nn
\eea
This integral can be calculated using the hypergeometric function identity \eqref{identity}. We get
\bea
(L_{\psi})_{\rom{in}} &\simeq&
4 \pi m_\psi Q_1 Q_5 C  e^{2\omega_I t}
\left(\omega_R R + \frac{m}{k} m_\psi -\frac{ n}{k} m_\phi   \right) k^{ 4 N + 2 l + 4} \times \cr
& & \frac{1}{2 (2 N + k |\zeta| + l + 2)} \frac{\Gamma(1+k|\zeta|)^2 \ \Gamma(N+1) \ \Gamma(N+l+2)}{\Gamma(N+k |\zeta|+l +2) \ \Gamma(N+k |\zeta| +1)}.
\eea
Using the definition of $\xi$ from \eqref{xi} and the integer relations \eqref{integer_replacements}, we obtain a contribution that is exactly the opposite of \eqref{Lpsifinal},
\bea
(L_{\psi})_{\rom{in}}\simeq - 2 \pi m_\psi Q_1 Q_5 C  e^{2\omega_I t}  k^{ 4 N + 2 l + 3}\frac{\Gamma(1+k|\zeta|)^2 \ \Gamma(N+1) \ \Gamma(N+l+2)}{\Gamma(N+k |\zeta|+l +2) \ \Gamma(N+k |\zeta| +1)} .
\eea

At a technical level the analysis presented above is significantly more involved compared to that of \cite{Chowdhury:2008bd}, however various technical pieces precisely fit together to give exactly equal and opposite contributions to $L_\psi$ (and hence $L_\phi$) from the inner and asymptotic regions.

\subsection{Energy and linear momentum of the perturbation}
\label{sec:energy}
A similar set of considerations applies to energy and linear momentum along $y$. Let us start with linear momentum along $y$.
The conserved linear momentum associated to the scalar perturbation is
\be
P_y =\int_{t=\rom{const}} \sqrt{-g}  dr  d \cA \ T^t_{~y}.
\ee
Using the separation ansatz \eqref{Psi_main_text} in the scalar stress tensor \eqref{stress_tensor}, the linear momentum expression reduces to
\be
P_y = \frac{2\lambda}{R} \int \sqrt{-g} dr d\mathcal{A} \left( - g^{tt} \omega_R + g^{t\psi} m_\psi + g^{t\phi} m_\phi + g^{ty} \frac{\lambda}{R} \right) \Psi \Psi^*.
\label{Py_simp}
\ee
Since the integral involved is exactly what we discussed above, it  follows that the inner and asymptotic region wavefunctions give equal and opposite contributions to the linear momentum.

The conserved energy of the scalar field $\Psi$ is
\be
 H=-\int_{t=\rom{const}}  \sqrt{-g}  dr  d\mathcal{A }  \ T^t_{~t}\,.
\ee
It is convenient to write this expression as a part which involves the integral already computed for the angular momentum, plus a remainder which is a total derivative~\cite{Chowdhury:2008bd}. We denote these as the bulk and boundary terms respectively,
\begin{equation}
H=H_\rom{bulk}+H_\rom{bdy}\,,
\end{equation}
where
\bea \label{Hbulk}
H_\rom{bulk}  &=& -\int \sqrt{-g}  dr d\mathcal{A}
\Bigl[g^{tt}\partial_t\Psi\partial_t\Psi^* \Bigr]
-\frac{1}{2}\int  dr d\mathcal{A} 
\Bigl[ \Psi \partial_i \left(\sqrt{-g}g^{ij}\partial_j\Psi^*\right)
+ \Psi^* \partial_i \left(\sqrt{-g}g^{ij}\partial_j\Psi\right) \Bigr],
\nonumber
\eea
\bea \label{Hboundary}
H_\rom{bdy} &=& \frac{1}{2}\int dr d\mathcal{A} \, \partial_i
\Bigl[\sqrt{-g}g^{ij} \partial_j \left( \Psi  \Psi^* \right)
\Bigr] .
\eea
Using the equation of motion for the scalar
$\partial_{\mu}(\sqrt{-g}\partial^{\mu}\Psi) = 0$
and the ansatz \eqref{Psi_main_text},
the bulk term simplifies to
\bea
\label{Hbulk-2}
&& H_\rom{bulk} ~=~ -2  \omega_R  \int \sqrt{-g} dr d\mathcal{A} \left( g^{tt} \omega_R -g^{t\phi} m_{\phi}- g^{t\psi} m_{\psi} -g^{ty}  \frac{\lambda}{R} \right) \Psi\Psi^* \,.
\eea

We now apply the decomposition into $H_\rom{bulk}$ and $H_\rom{bdy}$ separately in the inner and asymptotic regions.
For this purpose, we approximate both the outer boundary of the inner region and the inner boundary of the outer region by the surface $r=(Q_1Q_5)^{\frac{1}{4}}$.
Since the integral involved in $H_\rom{bulk}$  is exactly the one discussed above, it follows that the inner and asymptotic region wavefunctions give equal
and opposite contributions to $H_\rom{bulk}$. We need  only be concerned with the boundary terms. The only non-zero boundary terms arise from the terms with radial derivatives. We have
\bea
 H_\rom{bdy}^\rom{in}  &=&
 \frac{1}{2} \int d\mathcal{A} \left[\sqrt{-g}g^{rr}\partial_r (\Psi \Psi^*)\right]\Bigg{|}^{r=(Q_1Q_5)^{\frac{1}{4}}}_{r=r_+} 
~=~
H_\rom{bdy}^\rom{in,  neck} +H_\rom{bdy}^{\mathrm{in}, r=r_+} \,, \\
 H_\rom{bdy}^\rom{out}  &=&
 \frac{1}{2} \int d\mathcal{A} \left[\sqrt{-g}g^{rr}\partial_r (\Psi \Psi^*)\right]\Bigg{|}_{r=(Q_1Q_5)^{\frac{1}{4}}}^{r=\infty}
~=~
H_\rom{bdy}^{\mathrm{out},  r=\infty} + H_\rom{bdy}^\rom{out,  neck}.
\eea
We observe that
\be
H_\rom{bdy}^\rom{in, neck} = - H_\rom{bdy}^\rom{out, neck}. \label{matchH}
\ee

Let us now estimate the various boundary terms. Firstly, for $H_\rom{bdy}^{\mathrm{in},r=r_+} $,
counting powers of $\rho$ we see that as $\rho \to 0$,
\be
\sqrt{-g}g^{rr}\partial_r (\Psi \Psi^*)
= \sqrt{-g} \left[g^{\rho \rho} \left(\frac{d r}{ d\rho} \right)  \partial_\rho  (\Psi \Psi^*) \right]\\
\sim \rho^2 \partial_\rho  (\Psi \Psi^*) \sim  \rho^{2 k |\zeta| + 1},
\ee
which vanishes at $\rho = 0$ (i.e.~at $r=r_+$) and so we have $H_\rom{bdy}^{\mathrm{in},r=r_+} =0$.

Next, for $H_\rom{bdy}^{\mathrm{out},r=\infty} $  we observe that $\Psi \Psi^*$ falls off exponentially with exponent \eqref{power} in the $r \to \infty$ limit. Therefore, in the limit $r \to \infty$ it also vanishes.

Since we have observed in \eqref{matchH} that the neck terms are equal and opposite, it is not necessary to evaluate them to conclude that the contribution to the energy from the asymptotic region and the inner region are equal and opposite. Nevertheless, out of interest we now observe that these terms are parametrically subleading with respect to the contributions from $H_{\rom{bulk}}$.

At the neck, we have $\rho \sim \epsilon^{-1}$, so for $H_\rom{bdy}^{\mathrm{in, neck}} $ we find the parametric dependence
 \bea
 H_\rom{bdy}^{\mathrm{in, neck}}  &\sim& R\,\sqrt{-g} \left[g^{\rho \rho} \left(\frac{d r}{ d\rho} \right)  \partial_\rho  (\Psi \Psi^*)\right]\Bigg{|}_{\rho=1/\epsilon} \cr
 & \sim  &  R \, (Q_1 Q_5)^{\frac{3}{4}}\cdot \frac{\rho^2}{(Q_1 Q_5)^{\frac{1}{2}}} \cdot \frac{\rho}{(Q_1 Q_5)^{\frac{1}{4}}}  \frac{Q_1 Q_5}{R^2} \cdot \partial_\rho  \rho^{-2l -4}\Bigg{|}_{\rho=1/\epsilon}
~\,\sim~\,
(Q_1 Q_5)^{\frac{3}{4}} \epsilon^{2 l + 3}\,, \qquad\quad
 \label{estimate_in}
 \eea
and therefore $H_\rom{bdy}^{\mathrm{in, neck}} $ is subleading with respect to $H^{\mathrm{in}}_{\rom{bulk}}$.

Again it is not necessary to separately estimate $H_\rom{bdy}^{\mathrm{out, neck}}$, however it is straightforward to observe that as a result of the matching of solutions at the neck, the asymptotic region wavefunction also behaves as $\rho^{-l-2}$ in the neck, and with the same coefficient as the inner solution, giving precisely \eqref{estimate_in}.

To summarize, we have seen explicitly that the inner and asymptotic region wavefunctions give equal and opposite contributions to the conserved angular momenta, linear momentum along $y$ and energy of the scalar field. Since the inner part of the wavefunction carries negative energy with respect to the Killing vector $\partial_t$, it has its main support in the ergoregion. One can also see this fact explicitly by plotting a selection of examples.
Thus we see explicitly in this setup the physical picture of ergoregion emission as pair creation.


\section{Discussion}
\label{sec:discussion}

In this paper we have proposed the holographic description of the general family of orbifolded JMaRT solutions, with orbifolding parameter $k$. The $k>1$ states are of significant physical interest since states with larger $k$ are closer to typical states than states with smaller $k$. We have proposed that the dual CFT states are the general set of R-R and NS-NS states obtained by fractional spectral flow in both left- and right-moving sectors from the twisted NS-NS vacuum $|0_k\rangle_{\rom{NS}}$. 
We reviewed the fact that the orbifolded JMaRT solutions are completely smooth when the integer parameters $m$, $n$, and $k$ have no common divisors, and presented a full analysis of the orbifold singularities which arise depending on the common divisors between these parameters.

To support our proposed identification, we matched the minimal scalar emission spectrum and emission rate between gravity and CFT. On the gravity side, this involved solving the wave equation on the general orbifolded solution, generalizing previous studies~\cite{Cardoso:2005gj,Chowdhury:2008uj}. On the CFT side, our results were obtained via a straightforward generalization of the results of~\cite{Avery:2009xr}.

We also investigated the physical picture of ergoregion emission as pair creation, generalizing the results of \cite{Chowdhury:2008bd} to include three non-zero charges, two non-zero angular momenta, the orbifolding parameter $k$, and the most general form of the probe scalar wavefunction. 
We showed that radiation from the general orbifolded JMaRT solutions can be split into two distinct parts, one escaping to infinity and the other remaining deep inside the AdS region. 
Since the inner part of the wavefunction carries negative energy with respect to the Killing vector $\partial_t$ which generates time translations at spatial infinity, it has its main support in the ergoregion.

The states we have studied are non-BPS, and there is no known non-renormalization theorem protecting the quantities we have studied. Thus the fact that the orbifold CFT and gravity calculations agree exactly is quite non-trivial and better than might have been naively expected of the orbifold CFT. Naturally, the agreement observed in the $k=1$ solutions was reason for optimism on this point. The fact that our proposed dual states are related to BPS states by fractional spectral flow may perhaps be the feature which enables this non-trivial agreement.	

It would be interesting to study string theory in the subset of these backgrounds that have orbifold singularities.
String theory on orbifolds of AdS$_3 \times \rom{S}^3$ has previously been studied in~\cite{Martinec:2001cf,Martinec:2002xq}. In the presence of orbifold singularities, twisted sectors of closed strings typically give rise to light (or tachyonic) degrees of freedom that are not taken into account by supergravity~\cite{Dixon:1985jw,*Dixon:1986jc,*Dixon:1986qv}. Furthermore, non-supersymmetric orbifolds are expected to decay to a region of smooth spacetime together with an expanding pulse of excitations~\cite{Adams:2001sv,Gutperle:2002ki} (see also \cite{Vafa:2001ra,Harvey:2001wm})\footnote{One of the authors (DT) thanks Emil Martinec for a discussion on this point.}. 
If such a mechanism is present here, one can ask whether it interacts with the pair creation mechanism; for example one might imagine that the pair creation excitation that remains deep in the cap might interact with the orbifold and/or its decay products.

Indeed one might wonder whether such additional modes could affect the matching of emission frequencies and rates between the supergravity and CFT for the solutions with orbifold singularities. We did not find any such discrepancy; the calculations agree exactly between gravity and CFT regardless of the presence or absence of orbifold singularities. This strongly suggests that the ergoregion emission spectrum and rates are unaffected by such light degrees of freedom. 
It would be interesting to investigate this physics in more detail in the future.

\subsection*{Acknowledgements}
We thank Guillaume Bossard, Borun Chowdhury, Oleg Lunin, Emil Martinec, Samir Mathur, Kostas Skenderis, Marika Taylor and Nick Warner for fruitful discussions.  The work of DT was supported by the CEA Eurotalents program and by John Templeton Foundation Grant 48222: ``String Theory and the Anthropic Universe''. 
BC thanks TIFR for hospitality where part of this work was done. 
DT thanks AEI Potsdam for hospitality where part of this work was carried out.
AV thanks IPhT Saclay and IISER Bhopal for hospitality where part of this work was completed.



\begin{appendix}

\section[Appendices]{Solving the wave equation via matched asymptotic expansion}
\label{app:wave}

In this appendix we solve the wave equation in a matched asymptotic expansion analysis.
We obtain the instability frequencies and also fix the normalization of the wavefunction in the
asymptotic region given its form in the inner region.

We define the following regions of the geometry, in which we set up the matched asymptotic expansion. In Section \ref{sec:largeR} we specified that when studying AdS/CFT on the JMaRT solutions, one works in the regime of parameters
\bea
\epsilon = \frac{(Q_1 Q_5)^\frac{1}{4}}{R} ~  \ll ~ 1.
\label{epsilon-2}
\eea
In terms of the dimensionless radial variable $x$ defined in Eq.\;\eq{eq:x-def}, we define the `inner region' to be the range $ 0 \le  x \ll  \epsilon^{-2} $; to be more specific, let us introduce another parameter $\delta \ll 1$ and define the inner region to be given by\footnote{While one must consider $\epsilon$ to be exponentially small in order to get a large AdS inner region (see for example the discussion in \cite{Mathur:2011gz}), here $\delta$ is simply a bookkeeping device. The important point is that the inner and asymptotic regions do not overlap, and must be matched onto the neck region.}
\bea
0 \le  x \lesssim  \delta\frac{1}{\epsilon^{2}} \,.
\eea
We then define the `asymptotic region' to be given by the range
\bea
x \gtrsim \frac{1}{\delta}\frac{1}{\epsilon^{2}} \,.
\eea
The inner and asymptotic regions do not overlap. We will match solutions in the `neck' region $x \sim \frac{1}{\epsilon^{2}}$, or more specifically
\bea
\delta\frac{1}{\epsilon^{2}} \lesssim x \lesssim \frac{1}{\delta}\frac{1}{\epsilon^{2}}  
\eea
where  solutions to the radial wave equation are power law in $x$ \cite{Cvetic:1997uw}.  Solutions from the inner and asymptotic regions match on to these power law solutions from the two sides.\footnote{Note that the regions involved in the present matched asymptotic expansion analysis are different to those employed, e.g.,~in~\cite{Mathur:2011gz}.}

\subsubsection*{Inner region}

In the inner region one can neglect $\kappa^2 x$ relative to the other terms, and so
the radial wave equation \eqref{radial_eq} simplifies to
\be
4 \partial_x \left[x\left(x+\frac{1}{k^2}\right) \partial_x h\right] + \left(1- \nu^2 + \frac{\xi^2}{x + k^{-2}} - \frac{\zeta^2}{x} \right)h = 0.
\ee
Demanding regularity at the origin we get the solution for this equation
\be
h = \left(x + \frac{1}{k^2}\right)^{\frac{k\xi}{2}} x^{\frac{k|\zeta|}{2}} \left[{}_2F_1(a,b, c,- k^2 x)\right], \label{hinner}
\ee
where
\begin{align}
a &= \frac{1}{2} (1+ \nu + k|\zeta| + k\xi),
&b &= \frac{1}{2}(1 -\nu + k|\zeta| + k \xi),
&c &= 1+ k |\zeta|.
\end{align}

In writing this solution we have chosen to normalize the wavefunction \eqref{hinner} by setting its overall normalization constant to unity.
The behaviour of the inner solution near $x\to 0$ is simply $h \sim k^{-k \xi}x^\frac{k|\zeta|}{2},$ and its expansion for large $x$ is
\bea
h &\simeq& \Gamma(1+ k |\zeta|)
\Big{[}
\frac{k^{-1 - \nu - k |\zeta| - k \xi}\Gamma(-\nu)}{\Gamma\left(\frac{1}{2}(1 - \nu + k|\zeta| + k\xi)\right)\Gamma\left(\frac{1}{2}(1- \nu + k |\zeta| - k\xi))\right)} x^{-\frac{\nu +1}{2}} \nn
\\
& & \qquad \qquad +  \frac{k^{-1 + \nu - k |\zeta| - k \xi}\Gamma(\nu)}{\Gamma\left(\frac{1}{2}(1+  \nu + k |\zeta| + k \xi )\right)\Gamma\left(\frac{1}{2}(1+ \nu + k |\zeta| - k\xi)\right)} x^{\frac{\nu -1}{2}} \Big{]}.
\label{inner_expansion}
\eea
We will match this onto the power law behaviour in the neck region below.

\subsubsection*{Asymptotic region}

In the asymptotic region, one can neglect $\frac{\xi^2}{x + k^{-2}} - \frac{\zeta^2}{x} $ relative to the other terms, and so
the radial wave equation simplifies to
\be
\partial_x^2 (x h) + \left[ \frac{\kappa^2}{4x} + \frac{1-\nu^2}{4x^2} \right] (xh) =0.
\ee
The most general solution to this equation is a linear combination of Bessel functions
\be
h = \frac{1}{\sqrt{x}}\left[ C_1 J_\nu(\kappa \sqrt{x}) + C_2 J_{-\nu}(\kappa \sqrt{x}) \right].
\ee
For $\kappa \sqrt{x}  \ll 1$, its behaviour is
\be
h \sim \frac{C_1}{\Gamma(1+\nu)} \left(\frac{\kappa}{2}\right)^\nu x^{\frac{\nu-1}{2}} +   \frac{C_2}{\Gamma(1-\nu)} \left(\frac{\kappa}{2}\right)^{-\nu} x^{-\frac{\nu+1}{2}},
\label{outer_expansion}
\ee
and its large $\kappa \sqrt{x}$  behaviour is
\be
h \sim \frac{1}{x^\frac{3}{4}} \frac{1}{\sqrt{2\pi \kappa}} \left[ e^{i \kappa \sqrt{x}} e^{-i \frac{\pi}{4}}(C_1 e^{-i \nu \frac{\pi}{2}} + C_2 e^{i \nu \frac{\pi}{2}})
+  e^{-i \kappa \sqrt{x}}e^{i \frac{\pi}{4}} (C_1 e^{i \nu \frac{\pi}{2}} + C_2 e^{-i \nu \frac{\pi}{2}})\right]. \label{outer_wavefunction}
\ee

\subsubsection*{Neck region}
In the neck region, both $\kappa^2 x$ and $\frac{\xi^2}{x + k^{-2}} - \frac{\zeta^2}{x} $ can be neglected, and the wave equation approximates to
\be
\partial_x^2 (x h) + \left[ \frac{1-\nu^2}{4x^2}\right] (xh) =0.
\ee
The general solution is
\be
h = A \, x^{\frac{\nu-1}{2}} + B \, x^{-\frac{\nu+1}{2}}.
\label{eq:neck-sol}
\ee

\subsubsection*{Matching the solutions}
We can now match the solutions at each end of the neck region, and thereby patch together the three matching regions.

We are interested in instability of the geometry where there are no incoming waves, yet we have outgoing waves carrying energy and other charges to infinity. The
requirement of no incoming waves gives the relation
\be
C_1  + C_2 e^{-i \nu \pi} = 0. \label{outgoing}
\ee
Matching the two asymptotic expansions  \eqref{inner_expansion}  and \eqref{outer_expansion} to the solutions in the neck region, we obtain
\be
-e^{-i \pi \nu} \frac{\Gamma(1-\nu)}{\Gamma(1+\nu)} \left(\frac{\kappa}{2 k}\right)^{2\nu} =   \frac{\Gamma(\nu)}{\Gamma(-\nu)}
\frac{\Gamma\left(\frac{1}{2}(1-\nu + k |\zeta| + k \xi)\right)
\Gamma\left(\frac{1}{2}(1-\nu + k |\zeta| - k \xi)\right)}
{\Gamma\left(\frac{1}{2}(1+\nu + k |\zeta| + k \xi)\right)
\Gamma\left(\frac{1}{2}(1+\nu + k |\zeta| - k \xi)\right)}.
\label{matching}
\ee
The emission frequencies are given by the solutions to this transcendental equation.

\subsubsection*{Instability frequencies}
Let us now analyze equation \eqref{matching}. Recall that we work in the large $R$ limit, $\epsilon \ll 1$. In this limit, taking $\omega\sim 1/R$ and $\lambda \sim 1$, one finds $\kappa^2 \sim \epsilon^4$,
as can be seen from Eqs.~\eqref{kappa}, \eqref{rplusrminus}, and \eqref{epsilon}. Therefore the LHS of equation  \eqref{matching} is parametrically small. The RHS is parametrically small when one of the $\Gamma$ functions in the denominator is parametrically close to developing a pole. To leading order, the values of the parameters will be those which give poles.
Let us set\footnote{Taking parameters for which the other Gamma function in the denominator of \eq{matching} develops a pole leads to an exponentially decaying mode, rather than an exponentially growing mode.}
\begin{equation}
\frac{1}{2}(1+\nu + k |\zeta| + k \xi) \simeq -N,
\label{pole}
\end{equation}
with $N$ a non-negative integer. From equations $\eqref{varrho}$ and \eqref{vartheta}
we see that in the large $R$ limit $\varrho \to 1$
and $\vartheta \sim \epsilon^2$. Hence to leading order one obtains
\bea
\xi &\simeq& \omega R - m_\phi \frac{n}{k} + m_\psi \frac{m}{k},  \label{xilargeR-2}\\
\nu &\simeq& l + 1.
\eea
Replacing these relations in equation \eqref{pole} gives the leading order instability frequencies. To leading order, the instability frequencies are real; we define $\omega_R$ to be the real part of $\omega$, thus obtaining
\bea
\omega_R &\simeq&
\frac{1}{k R } \Bigl( - l - m_\psi m  + m_\phi n  -  \left|-  k \lambda - m_\psi n + m_\phi m \right| -   2(N + 1) \Bigr).
\eea

At next-to-leading order, we will obtain the leading imaginary part of $\omega$.
To do this, we replace $ N \to N + \delta N$ and eliminate $\xi$ in favour of $N$.
From equation \eqref{xilargeR} we have $\delta \xi =R \delta \omega$, which upon using \eqref{pole} gives the change in $\omega$ due to shifting $N$ to be
\bea
\delta \omega &=& -\frac{2}{k R}\delta N \,.
\eea
There are also contributions to the subleading part of $\omega$ from corrections to $\nu$ and $\xi$ at order $\epsilon^2$, however these affect only the real part of $\omega$.
Therefore, denoting by $\omega_I$ the imaginary part of $\omega$, to leading order in $\epsilon$ we have
\be
\omega_I \simeq -\frac{2}{k R} \mbox{Im}(\delta N).
\ee

The small deformation $\delta N$ controls the pole of the divergent $\Gamma$ function. We assume that $\delta N \ll \epsilon$, so that to leading order in $\epsilon$ it can be neglected in the argument of all the other $\Gamma$ functions.
In what follows we shall verify the consistency of this assumption.
The residue at the pole of the $\Gamma$ function is given by
\be
\Gamma(-N-\delta N)=\frac{(-1)^{N+1}}{N!}\frac{1}{\delta N}.
\ee
Using the relations
\begin{align}
\Gamma(n+1+x) &= x n! [x]_n \Gamma(x),  &
\Gamma(-n-x) &= \frac{\Gamma(-x)}{(-1)^n n! [x]_n}, &
\end{align}
where $[x]_n=\prod\limits_{i=1}^{n} \left(1+\frac{x}{i}\right),$  we obtain
\be
\delta N=-e^{-i \pi \nu} \left(\frac{\nu \Gamma(-\nu)}{\Gamma(\nu)}\right) \left( \frac{\kappa}{2k}\right)^{2\nu} [\nu]_N  [\nu]_{N+ k |{\zeta}|}.
\label{deltaN}
\ee
For $p$ and $q$ integers, we have $[p]_q ={}^{p+q}C_{p} = {p+q \choose p}$.

The identity
\begin{equation}
\Gamma(\nu) \Gamma(-\nu)= \frac{-\pi}{\nu \sin(\pi\nu)}
\end{equation}
allows us to extract $\mbox{Im}(\delta N)$. From \eqref{deltaN} we obtain
\be
\omega_I \simeq \frac{2}{k R} \frac{\pi}{\Gamma(\nu)^2} \left( \frac{\kappa}{2k} \right)^{2\nu} [\nu]_N  [\nu]_{N+ k |{\zeta}|} \,.
\label{omegaI}
\ee
Recalling that $\nu=l+1+\cO(\epsilon^2)$, we observe that $\mbox{Im} (\delta N) \sim \epsilon^{4l+4}$, and that $\mbox{Re} (\delta N) \sim \epsilon^{4l+2}$, which demonstrates the consistency of our approach.
Then to leading order the imaginary part of the frequency is
\bea
\omega_I &\simeq&
\frac{1} {kR} \frac{\pi}{2^{2 l + 1}(l!)^2}
\left[\left(\omega^2-\frac{\lambda^2}{R^2}\right) \frac{Q_1 Q_5}{k^2 R^2} \right]^{l+1}
{N + l + 1\choose l+1} {N+k| \zeta | + l + 1 \choose l+1} \,.  \qquad
\label{omegaIFinal}
\eea


\refstepcounter{section}
\section*{\thesection \quad Details of pair creation calculation}
\label{app:pair_details}

\refstepcounter{subsection}
\subsection*{\thesubsection ~ Normalization of the asymptotic region wavefunction}

In this appendix we fix the normalization of the asymptotic region wavefunction, for use in Section \ref{pair_creation}.

Using the asymptotic region wavefunction \eqref{outer_wavefunction} together with the requirement of only outgoing waves \eqref{outgoing}, we obtain
\be \label{eq:asym-h-outgoing}
h_\rom{out}(x) = C_2  \frac{1}{\sqrt{2\pi \kappa}} \frac{1}{x^{\frac{3}{4}}}e^{i \kappa \sqrt{x}} e^{-i \frac{\pi}{4}}  \left(e^{i \frac{\pi \nu}{2}} - e^{-i \frac{3\pi \nu}{2}}\right),
\ee
and thus
\be
h_\rom{out}(x) h^*_\rom{out}(x) = |C_2|^2  \frac{2}{\pi |\kappa|} \frac{1}{x^{\frac{3}{2}}}e^{i (\kappa -\kappa^*) \sqrt{x}}  \sin^2 (\pi \nu).
\label{hhstar}
\ee
Matching the two asymptotic expansions -- \eqref{outer_expansion} and \eqref{inner_expansion} -- say by comparing coefficients of $x^{\frac{\nu-1}{2}}$, we get an equation that determines $C_2$,
\begin{equation}
\frac{k^{-1 + \nu - k |\zeta| - k \xi}\Gamma(1+ k |\zeta|) \Gamma(\nu)}{\Gamma\left(\frac{1}{2}(1+ \nu + k|\zeta| + k\xi)\right)\Gamma\left(\frac{1}{2}(1+ \nu + k |\zeta| - k \xi)\right)} = (-C_2 e^{-i \pi \nu}) \frac{1}{\Gamma(1+\nu)} \left(\frac{\kappa}{2}\right)^\nu.
\label{c2equation}
\end{equation}
To find the real and imaginary frequencies we matched the solution using
\be
\Gamma\left(\frac{1}{2}(1+ \nu + k |\zeta| + k \xi)\right) = \Gamma(- N - \delta N) = \frac{(-1)^{N+1}}{N! \delta N}.
\ee
Replacing this expression in \eqref{c2equation} we get
\begin{equation}
\frac{k^{2N + 2 \nu}\Gamma(1+ k|\zeta|) \Gamma(\nu)}{\Gamma(N  +  \nu + 1 + k|\zeta|)} (-1)^{N+1} N! \delta N
= (- C_2 e^{-i\pi \nu})  \frac{1}{\Gamma(1+\nu)} \left(\frac{\kappa}{2}\right)^\nu.
\end{equation}
Taking modulus of the above relation allows us to extract $|C_2|^2$. We get,
\begin{equation}
\frac{k^{4N + 4 \nu}\Gamma(1+k |\zeta|)^2 \Gamma(\nu)^2 \Gamma(N+1)^2}{\Gamma(N  + \nu + 1+ k |\zeta|)^2}  (\delta N)(\delta N)^*
= |C_2|^2 \frac{1}{\Gamma(1+\nu)^2} \left(\frac{|\kappa|}{2}\right)^{2\nu}.
\label{normc2eq}
\end{equation}
We now use \eqref{deltaN}, and working to leading order in $\epsilon$, we approximate $|\kappa|^2 \simeq \kappa^2$.
For use in the main text, it is convenient to extract one power of $\omega_I$ using \eq{omegaI}. We thus obtain
\be
|C_2|^2 \sin^2(\pi\nu) = \frac{ \pi}{2} k^{4 N + 2 \nu}(k R)\omega_I \,
\frac{\Gamma(1+k|\zeta|)^2  \  \Gamma(N+1) \ \Gamma(N+\nu+1)}{\Gamma(N+\nu +1+k|\zeta|) \ \Gamma(N+1+k|\zeta|)} \,.
\label{normc2}
\ee

\refstepcounter{subsection}
\subsection*{\thesubsection ~ A hypergeometric function identity}

{\bf Identity:} For positive $\gamma$ and for arbitrary positive integers $N$ and $l$,
\bea
&& \int_{0}^{\infty}  d \rho \rho^{2 \gamma +1} (1 + \rho^2)^{-2N- l -3 -\gamma}   ({}_2F_1(-N,-N -l-1,1+\gamma,-\rho^2))^2 \nn \\
&& = \frac{1}{2 (2 N + \gamma + l + 2)} \frac{\Gamma(1+\gamma)^2 \ \Gamma(N+1) \ \Gamma(N+l+2)}{\Gamma(N+\gamma+l +2) \ \Gamma(N+\gamma+1)}.
\label{identity}
\eea
{\bf Proof:} A proof of the above identity can be given by relating hypergeometric functions in the integral to Jacobi polynomials. From identity 8.962.1 (third line) of Gradshteyn and Ryzhik \cite{gradshteyn2007}, page  999, we have
\be
P^{(\gamma, l +1)}_N (y) = \frac{\Gamma(N + 1 + \gamma)}{\Gamma(N+1)\Gamma(1+\gamma)}\left(\frac{1+y}{2}\right)^N {}_2F_1\left(-N,-N -l-1,1+\gamma,\frac{y-1}{y+1}\right).
\ee
Defining
\be
\frac{y-1}{y+1} = - \rho^2,
\ee
the integral can be converted into
\be
\frac{1}{2^{\gamma + l + 3}}\left( \frac{\Gamma(N+1)\Gamma(1+\gamma)}{\Gamma(N + 1 + \gamma)}\right)^2 \int_{-1}^{1} dy (1-y)^{\gamma}(1+y)^{l+1} \left(P^{(\gamma, l +1)}_N (y)\right)^2,
\ee
which simply gives the right hand side of \eqref{identity} upon using  identity 7.391.1 (second line) on page  806 of \cite{gradshteyn2007}.

\newpage

\refstepcounter{section}
\section*{\thesection \quad Conventions} \label{app:conventions}

In this appendix we record our conventions and their relation to those of Ref.~\cite{Avery:2009xr}, which we use to obtain Eq.~\eq{eq:cftspectrum1} of the main text. 

Our conventions are that
\be
\mathrm{left}\mhyphen\mathrm{moving} ~~\leftrightarrow~~ \mathrm{holomorphic} ~~\leftrightarrow~~
\mathrm{positive}~ P_y,
\ee
where $P_y$ is momentum along $y$. So the holomorphic coordinate in the CFT is related to the null coordinate $v=(t-y)$ in the spacetime.

Our map between CFT and gravity SU(2) quantum numbers is given in~\eq{eq:cftSU2s},
\bea
m_\psi &=& -(m+\bar m)  \,, \qquad
 m_\phi ~=~  (m-\bar m) \,.
\eea

Let us compare our conventions to those of Avery-Chowdhury \cite{Avery:2009xr}, whose quantities we denote with a superscript AC. In that paper, the anti-holomorphic coordinate corresponds to positive $y$. Therefore we interchange $L$ and $R$ in mapping between the two papers, so the spectral flow parameters are
\bea
\alpha &=& \bar \alpha^{AC} \,, \qquad 
\bar{\alpha} ~=~ \alpha^{AC}\,. 
\label{eq:alphamap}
\eea
Next, the parameter controlling the twist is
\bea
\kappa^{AC} &=& k \,.
\eea

In the conventions of \cite{Avery:2009xr}, the emission of a scalar with gravity quantum numbers $(l,m_\psi^{AC},m_\phi^{AC})$ corresponds to the CFT vertex
\be
\cV_{l,-m_{\psi}^{AC},-m_{\phi}^{AC}}
\ee
where
\bea
-m_{\psi}^{AC} &=& l-k^{AC}-\bar k^{AC} \,, \qquad\quad
-m_{\phi}^{AC} ~=~ k^{AC}-\bar k^{AC} \,.
\eea
In addition, similarly to our conventions we have the relation
\bea
m_\psi^{AC} &=& -(m^{AC} + \bar m^{AC})\,, \qquad\quad
m_\phi^{AC} ~=~ m^{AC} - \bar m^{AC} \,.
\eea
Since $L$ and $R$ are interchanged between the two papers, we have
\bea
m_L &=& \bar m^{AC} \,, \qquad 
m_R ~=~ m^{AC}\, \qquad \Rightarrow \qquad 
m_\psi~=~ m_\psi^{AC} \,, \qquad 
m_\phi ~=~ - m_\phi^{AC}\,.
\eea
Therefore we obtain
\bea
k^{AC} &=& \frac12 \left( l + m_\psi + m_\phi \right)  \,, \qquad\quad 
\bar k^{AC} ~=~ \frac12 \left( l + m_\psi - m_\phi \right) \,.
\eea
Using these relations in Eq.\;(10.3) of \cite{Avery:2009xr}, we arrive at~\eq{eq:cftspectrum1}.

\end{appendix}

\newpage

\begin{adjustwidth}{-3mm}{-3mm} 

\bibliography{JMaRT}
\bibliographystyle{utphysM}

\end{adjustwidth}

\end{document}